\ifx\newheadisloaded\relax\immediate\write16{***already loaded} \else\let\newheadisloaded=\relax\fi
\gdef\isonnarrowscreen{F}
\gdef\PSfonts{T}
\magnification\magstep1

\newdimen\papwidth
\newdimen\papheight
\newskip\beforesectionskipamount  
\newskip\sectionskipamount 
\def\sectionskip{\vskip\sectionskipamount}
\def\beforesectionskip{\vskip\beforesectionskipamount}
\papwidth=16truecm
\if F\isonnarrowscreen
\papheight=22truecm
\voffset=0.4truecm
\hoffset=0.4truecm
\else
\papheight=22truecm
\voffset=-1.5truecm
\hoffset=-1truecm
\fi
\hsize=\papwidth
\vsize=\papheight
\catcode`\@=11
\ifx\amstexloaded@\relax
\else
\nopagenumbers
\headline={\ifnum\pageno>1 {\hss\tenrm-\ \folio\ -\hss} \else
{\hfill}\fi}
\fi
\catcode`\@=\active
\newdimen\texpscorrection
\texpscorrection=0.15truecm 

\def\sectionsize{\twelvepoint}
\def\sectiontype{\bf}
\def\subsectionsize{}
\def\subsectiontype{\bf}
\def\em{\sl}
\newfam\truecmsy
\newfam\truecmr%
\newfam\msafam%
\newfam\msbfam
\newfam\scriptfam
\newfam\frakfam
\newfam\frakbfam
\newskip\ttglue 
\if T\isonnarrowscreen
\papheight=11.5truecm
\fi
\if F\PSfonts
\font\twelverm=cmr12
\font\tenrm=cmr10
\font\eightrm=cmr8
\font\sevenrm=cmr7
\font\sixrm=cmr6
\font\fiverm=cmr5

\font\twelvebf=cmbx12
\font\tenbf=cmbx10
\font\eightbf=cmbx8
\font\sevenbf=cmbx7
\font\sixbf=cmbx6
\font\fivebf=cmbx5

\font\twelveit=cmti12
\font\tenit=cmti10
\font\eightit=cmti8
\font\sevenit=cmti7
\font\sixit=cmti6
\font\fiveit=cmti5

\font\twelvesl=cmsl12
\font\tensl=cmsl10
\font\eightsl=cmsl8
\font\sevensl=cmsl7
\font\sixsl=cmsl6
\font\fivesl=cmsl5

\font\twelvei=cmmi12
\font\teni=cmmi10
\font\eighti=cmmi8
\font\seveni=cmmi7
\font\sixi=cmmi6
\font\fivei=cmmi5

\font\twelvesy=cmsy10	at	12pt
\font\tensy=cmsy10
\font\eightsy=cmsy8
\font\sevensy=cmsy7
\font\sixsy=cmsy6
\font\fivesy=cmsy5
\font\twelvetruecmsy=cmsy10	at	12pt
\font\tentruecmsy=cmsy10
\font\eighttruecmsy=cmsy8
\font\seventruecmsy=cmsy7
\font\sixtruecmsy=cmsy6
\font\fivetruecmsy=cmsy5

\font\twelvetruecmr=cmr12
\font\tentruecmr=cmr10
\font\eighttruecmr=cmr8
\font\seventruecmr=cmr7
\font\sixtruecmr=cmr6
\font\fivetruecmr=cmr5

\font\twelvebf=cmbx12
\font\tenbf=cmbx10
\font\eightbf=cmbx8
\font\sevenbf=cmbx7
\font\sixbf=cmbx6
\font\fivebf=cmbx5

\font\twelvett=cmtt12
\font\tentt=cmtt10
\font\eighttt=cmtt8

\font\twelveex=cmex10	at	12pt
\font\tenex=cmex10

\font\twelvemsb=msbm10	at	12pt
\font\tenmsb=msbm10
\font\eightmsb=msbm8
\font\sevenmsb=msbm7
\font\sixmsb=msbm6
\font\fivemsb=msbm5

\font\twelvemsa=msam10	at	12pt
\font\tenmsa=msam10
\font\eightmsa=msam8
\font\sevenmsa=msam7
\font\sixmsa=msam6
\font\fivemsa=msam5


\font\tenfrm=eufm10
\font\eightfrm=eufm8
\font\sevenfrm=eufm7
\font\sixfrm=eufm6
\font\fivefrm=eufm5

\font\tenfrb=eufb10
\font\eightfrb=eufb8
\font\sevenfrb=eufb7
\font\sixfrb=eufb6
\font\fivefrb=eufb5
\font\twelvescr=eusm10 at 12pt
\font\tenscr=eusm10
\font\eightscr=eusm8
\font\sevenscr=eusm7
\font\sixscr=eusm6
\font\fivescr=eusm5
\fi
\if T\PSfonts
\font\twelverm=ptmr	at	12pt
\font\tenrm=ptmr	at	10pt
\font\eightrm=ptmr	at	8pt
\font\sevenrm=ptmr	at	7pt
\font\sixrm=ptmr	at	6pt
\font\fiverm=ptmr	at	5pt

\font\twelvebf=ptmb	at	12pt
\font\tenbf=ptmb	at	10pt
\font\eightbf=ptmb	at	8pt
\font\sevenbf=ptmb	at	7pt
\font\sixbf=ptmb	at	6pt
\font\fivebf=ptmb	at	5pt

\font\twelveit=ptmri	at	12pt
\font\tenit=ptmri	at	10pt
\font\eightit=ptmri	at	8pt
\font\sevenit=ptmri	at	7pt
\font\sixit=ptmri	at	6pt
\font\fiveit=ptmri	at	5pt

\font\twelvesl=ptmro	at	12pt
\font\tensl=ptmro	at	10pt
\font\eightsl=ptmro	at	8pt
\font\sevensl=ptmro	at	7pt
\font\sixsl=ptmro	at	6pt
\font\fivesl=ptmro	at	5pt

\font\twelvei=cmmi12
\font\teni=cmmi10
\font\eighti=cmmi8
\font\seveni=cmmi7
\font\sixi=cmmi6
\font\fivei=cmmi5

\font\twelvesy=cmsy10	at	12pt
\font\tensy=cmsy10
\font\eightsy=cmsy8
\font\sevensy=cmsy7
\font\sixsy=cmsy6
\font\fivesy=cmsy5
\font\twelvetruecmsy=cmsy10	at	12pt
\font\tentruecmsy=cmsy10
\font\eighttruecmsy=cmsy8
\font\seventruecmsy=cmsy7
\font\sixtruecmsy=cmsy6
\font\fivetruecmsy=cmsy5

\font\twelvetruecmr=cmr12
\font\tentruecmr=cmr10
\font\eighttruecmr=cmr8
\font\seventruecmr=cmr7
\font\sixtruecmr=cmr6
\font\fivetruecmr=cmr5


\font\twelvett=cmtt12
\font\tentt=cmtt10
\font\eighttt=cmtt8

\font\twelveex=cmex10	at	12pt
\font\tenex=cmex10

\font\twelvemsb=msbm10	at	12pt
\font\tenmsb=msbm10
\font\eightmsb=msbm8
\font\sevenmsb=msbm7
\font\sixmsb=msbm6
\font\fivemsb=msbm5
\font\twelvemsa=msam10	at	12pt
\font\tenmsa=msam10
\font\eightmsa=msam8
\font\sevenmsa=msam7
\font\sixmsa=msam6
\font\fivemsa=msam5


\font\tenfrm=eufm10
\font\eightfrm=eufm8
\font\sevenfrm=eufm7
\font\sixfrm=eufm6
\font\fivefrm=eufm5

\font\tenfrb=eufb10
\font\eightfrb=eufb8
\font\sevenfrb=eufb7
\font\sixfrb=eufb6
\font\fivefrb=eufb5
\font\twelvescr=eusm10 at 12pt
\font\tenscr=eusm10
\font\eightscr=eusm8
\font\sevenscr=eusm7
\font\sixscr=eusm6
\font\fivescr=eusm5
\fi
\def\eightpoint{\def\rm{\fam0\eightrm}%
\textfont0=\eightrm
  \scriptfont0=\sixrm
  \scriptscriptfont0=\fiverm 
\textfont1=\eighti
  \scriptfont1=\sixi
  \scriptscriptfont1=\fivei 
\textfont2=\eightsy
  \scriptfont2=\sixsy
  \scriptscriptfont2=\fivesy 
\textfont3=\tenex
  \scriptfont3=\tenex
  \scriptscriptfont3=\tenex 
\textfont\itfam=\eightit
  \scriptfont\itfam=\sixit
  \scriptscriptfont\itfam=\fiveit 
  \def\it{\fam\itfam\eightit}%
\textfont\slfam=\eightsl
  \scriptfont\slfam=\sixsl
  \scriptscriptfont\slfam=\fivesl 
  \def\sl{\fam\slfam\eightsl}%
\textfont\ttfam=\eighttt
  \def\tt{\fam\ttfam\eighttt}%
\textfont\bffam=\eightbf
  \scriptfont\bffam=\sixbf
  \scriptscriptfont\bffam=\fivebf
  \def\bf{\fam\bffam\eightbf}%
\textfont\frakfam=\eightfrm
  \scriptfont\frakfam=\sixfrm
  \scriptscriptfont\frakfam=\fivefrm
  \def\frak{\fam\frakfam\eightfrm}%
\textfont\frakbfam=\eightfrb
  \scriptfont\frakbfam=\sixfrb
  \scriptscriptfont\frakbfam=\fivefrb
  \def\bfrak{\fam\frakbfam\eightfrb}%
\textfont\scriptfam=\eightscr
  \scriptfont\scriptfam=\sixscr
  \scriptscriptfont\scriptfam=\fivescr
  \def\script{\fam\scriptfam\eightscr}%
\textfont\msbfam=\eightmsb
  \scriptfont\msbfam=\sixmsb
  \scriptscriptfont\msbfam=\fivemsb
  \def\bb{\fam\msbfam\eightmsb}%
\textfont\msafam=\eightmsa
  \scriptfont\msafam=\sixmsa
  \scriptscriptfont\msafam=\fivemsa
\textfont\truecmr=\eighttruecmr
  \scriptfont\truecmr=\sixtruecmr
  \scriptscriptfont\truecmr=\fivetruecmr
  \def\truerm{\fam\truecmr\eighttruecmr}%
\textfont\truecmsy=\eighttruecmsy
  \scriptfont\truecmsy=\sixtruecmsy
  \scriptscriptfont\truecmsy=\fivetruecmsy
\tt \ttglue=.5em plus.25em minus.15em 
\normalbaselineskip=9pt
\setbox\strutbox=\hbox{\vrule height7pt depth2pt width0pt}%
\normalbaselines
\rm
}

\def\tenpoint{\def\rm{\fam0\tenrm}%
\textfont0=\tenrm
  \scriptfont0=\sevenrm
  \scriptscriptfont0=\fiverm 
\textfont1=\teni
  \scriptfont1=\seveni
  \scriptscriptfont1=\fivei 
\textfont2=\tensy
  \scriptfont2=\sevensy
  \scriptscriptfont2=\fivesy 
\textfont3=\tenex
  \scriptfont3=\tenex
  \scriptscriptfont3=\tenex 
\textfont\itfam=\tenit
  \scriptfont\itfam=\sevenit
  \scriptscriptfont\itfam=\fiveit 
  \def\it{\fam\itfam\tenit}%
\textfont\slfam=\tensl
  \scriptfont\slfam=\sevensl
  \scriptscriptfont\slfam=\fivesl 
  \def\sl{\fam\slfam\tensl}%
\textfont\ttfam=\tentt
  \def\tt{\fam\ttfam\tentt}%
\textfont\bffam=\tenbf
  \scriptfont\bffam=\sevenbf
  \scriptscriptfont\bffam=\fivebf
  \def\bf{\fam\bffam\tenbf}%
\textfont\frakfam=\tenfrm
  \scriptfont\frakfam=\sevenfrm
  \scriptscriptfont\frakfam=\fivefrm
  \def\frak{\fam\frakfam\tenfrm}%
\textfont\frakbfam=\tenfrb
  \scriptfont\frakbfam=\sevenfrb
  \scriptscriptfont\frakbfam=\fivefrb
  \def\bfrak{\fam\frakbfam\tenfrb}%
\textfont\scriptfam=\tenscr
  \scriptfont\scriptfam=\sevenscr
  \scriptscriptfont\scriptfam=\fivescr
  \def\script{\fam\scriptfam\tenscr}%
\textfont\msbfam=\tenmsb
  \scriptfont\msbfam=\sevenmsb
  \scriptscriptfont\msbfam=\fivemsb
  \def\bb{\fam\msbfam\tenmsb}%
\textfont\msafam=\tenmsa
  \scriptfont\msafam=\sevenmsa
  \scriptscriptfont\msafam=\fivemsa%
\textfont\truecmr=\tentruecmr
  \scriptfont\truecmr=\seventruecmr
  \scriptscriptfont\truecmr=\fivetruecmr
  \def\truerm{\fam\truecmr\tentruecmr}%
\textfont\truecmsy=\tentruecmsy
  \scriptfont\truecmsy=\seventruecmsy
  \scriptscriptfont\truecmsy=\fivetruecmsy
\tt \ttglue=.5em plus.25em minus.15em 
\normalbaselineskip=12pt
\setbox\strutbox=\hbox{\vrule height8.5pt depth3.5pt width0pt}%
\normalbaselines
\rm
}

\def\twelvepoint{\def\rm{\fam0\twelverm}%
\textfont0=\twelverm
  \scriptfont0=\tenrm
  \scriptscriptfont0=\eightrm 
\textfont1=\twelvei
  \scriptfont1=\teni
  \scriptscriptfont1=\eighti 
\textfont2=\twelvesy
  \scriptfont2=\tensy
  \scriptscriptfont2=\eightsy 
\textfont3=\twelveex
  \scriptfont3=\twelveex
  \scriptscriptfont3=\twelveex 
\textfont\itfam=\twelveit
  \scriptfont\itfam=\tenit
  \scriptscriptfont\itfam=\eightit 
  \def\it{\fam\itfam\twelveit}%
\textfont\slfam=\twelvesl
  \scriptfont\slfam=\tensl
  \scriptscriptfont\slfam=\eightsl 
  \def\sl{\fam\slfam\twelvesl}%
\textfont\ttfam=\twelvett
  \def\tt{\fam\ttfam\twelvett}%
\textfont\bffam=\twelvebf
  \scriptfont\bffam=\tenbf
  \scriptscriptfont\bffam=\eightbf
  \def\bf{\fam\bffam\twelvebf}%
\textfont\scriptfam=\twelvescr
  \scriptfont\scriptfam=\tenscr
  \scriptscriptfont\scriptfam=\eightscr
  \def\script{\fam\scriptfam\twelvescr}%
\textfont\msbfam=\twelvemsb
  \scriptfont\msbfam=\tenmsb
  \scriptscriptfont\msbfam=\eightmsb
  \def\bb{\fam\msbfam\twelvemsb}%
\textfont\msafam=\twelvemsa
  \scriptfont\msafam=\tenmsa
  \scriptscriptfont\msafam=\eightmsa%
\textfont\truecmr=\twelvetruecmr
  \scriptfont\truecmr=\tentruecmr
  \scriptscriptfont\truecmr=\eighttruecmr
  \def\truerm{\fam\truecmr\twelvetruecmr}%
\textfont\truecmsy=\twelvetruecmsy
  \scriptfont\truecmsy=\tentruecmsy
  \scriptscriptfont\truecmsy=\eighttruecmsy
\tt \ttglue=.5em plus.25em minus.15em 
\setbox\strutbox=\hbox{\vrule height7pt depth2pt width0pt}%
\normalbaselineskip=15pt
\normalbaselines
\rm
}
%
\fontdimen16\tensy=2.7pt
\fontdimen13\tensy=4.3pt
\fontdimen17\tensy=2.7pt
\fontdimen14\tensy=4.3pt
\fontdimen18\tensy=4.3pt
\fontdimen16\eightsy=2.7pt
\fontdimen13\eightsy=4.3pt
\fontdimen17\eightsy=2.7pt
\fontdimen14\eightsy=4.3pt
\fontdimen18\sevensy=4.3pt
\fontdimen16\sevensy=1.8pt
\fontdimen13\sevensy=4.3pt
\fontdimen17\sevensy=2.7pt
\fontdimen14\sevensy=4.3pt
\fontdimen18\sevensy=4.3pt
%
\def\hexnumber#1{\ifcase#1 0\or1\or2\or3\or4\or5\or6\or7\or8\or9\or
 A\or B\or C\or D\or E\or F\fi}
\mathcode`\=="3\hexnumber\truecmr3D
\mathchardef\not="3\hexnumber\truecmsy36
\mathcode`\+="2\hexnumber\truecmr2B
\mathcode`\(="4\hexnumber\truecmr28
\mathcode`\)="5\hexnumber\truecmr29
\mathcode`\!="5\hexnumber\truecmr21
\mathcode`\(="4\hexnumber\truecmr28
\mathcode`\)="5\hexnumber\truecmr29

\def\tilde{\mathaccent"0\hexnumber\truecmr7E }
\def\bar{\mathaccent"0\hexnumber\truecmr16 }

\def\Phi{\mathchar"0\hexnumber\truecmr08 }
\def\Gamma {\mathchar"0\hexnumber\truecmr00 }
\def\Delta {\mathchar"0\hexnumber\truecmr01 }
\def\Theta {\mathchar"0\hexnumber\truecmr02 }
\def\Lambda{\mathchar"0\hexnumber\truecmr03 }
\def\Xi {\mathchar"0\hexnumber\truecmr04 }
\def\Pi{\mathchar"0\hexnumber\truecmr05 }
\def\Sigma{\mathchar"0\hexnumber\truecmr06 }
\def\Upsilon {\mathchar"0\hexnumber\truecmr07 }
\def\Phi {\mathchar"0\hexnumber\truecmr08 }
\def\Psi {\mathchar"0\hexnumber\truecmr09 }
\def\Omega{\mathchar"0\hexnumber\truecmr0A }
\newcount\EQNcount \EQNcount=1
\newcount\CLAIMcount \CLAIMcount=1
\newcount\SECTIONcount \SECTIONcount=0
\newcount\SUBSECTIONcount \SUBSECTIONcount=1
\def\ifff#1#2#3{\ifundefined{#1#2}%
\expandafter\xdef\csname #1#2\endcsname{#3}\else%
\fi}
\def\NEWDEF#1#2#3{\ifff{#1}{#2}{#3}}
\def\actualnumber{\number\SECTIONcount}
\def\EQ#1{\lmargin{#1}\eqno\tageck{#1}}
\def\tageck#1{\lmargin{#1}({\rm \actualnumber}.\number\EQNcount)
 \NEWDEF{e}{#1}{(\actualnumber.\number\EQNcount)}
\global\advance\EQNcount by 1
}
\def\SECT#1#2{\lmargin{#1}\SECTION{#2}%
\NEWDEF{s}{#1}{\actualnumber}%
}

\def\CLAIM#1#2#3\par{
\vskip.1in\medbreak\noindent
{\lmargin{#2}\bf #1\ \actualnumber.\number\CLAIMcount.} {\sl #3}\par
\NEWDEF{c}{#2}{#1\ \actualnumber.\number\CLAIMcount}
\global\advance\CLAIMcount by 1
\ifdim\lastskip<\medskipamount
\removelastskip\penalty55\medskip\fi}
\def\CLAIMNONR #1#2#3\par{
\vskip.1in\medbreak\noindent
{\lmargin{#2}\bf #1.} {\sl #3}\par
\NEWDEF{c}{#2}{#1}
\global\advance\CLAIMcount by 1
\ifdim\lastskip<\medskipamount
\removelastskip\penalty55\medskip\fi}
\def\SECTION#1{\vskip0pt plus.2\vsize\penalty-75
    \vskip0pt plus -.2\vsize
    \global\advance\SECTIONcount by 1
    \beforesectionskip\noindent
{\sectionsize\sectiontype \actualnumber.\ #1}
    \EQNcount=1
    \CLAIMcount=1
    \SUBSECTIONcount=1
    \nobreak\sectionskip\noindent}
\def\SECTIONNONR#1{\vskip0pt plus.3\vsize\penalty-75
    \vskip0pt plus -.3\vsize
    \global\advance\SECTIONcount by 1
    \beforesectionskip\noindent
{\sectionsize\sectiontype  #1}
     \EQNcount=1
     \CLAIMcount=1
     \SUBSECTIONcount=1
     \nobreak\sectionskip\noindent}
\def\SUBSECTION#1{\vskip0pt plus.2\vsize\penalty-75%
    \vskip0pt plus -.2\vsize%
    \beforesectionskip\noindent%
{\subsectionsize\subsectiontype \actualnumber.\number\SUBSECTIONcount.\ #1}
    \global\advance\SUBSECTIONcount by 1
    \nobreak\sectionskip\noindent}
\def\SUBSECTIONNONR#1\par{\vskip0pt plus.2\vsize\penalty-75
    \vskip0pt plus -.2\vsize
\beforesectionskip\noindent
{\subsectionsize\subsectiontype #1}
    \nobreak\sectionskip\noindent\noindent}
\def\ifundefined#1{\expandafter\ifx\csname#1\endcsname\relax}
\def\equ#1{\ifundefined{e#1}$\spadesuit$#1\else\csname e#1\endcsname\fi}
\def\clm#1{\ifundefined{c#1}$\spadesuit$#1\else\csname c#1\endcsname\fi}
\def\sec#1{\ifundefined{s#1}$\spadesuit$#1
\else Section \csname s#1\endcsname\fi}
\def\lab#1#2{\ifundefined{#1#2}$\spadesuit$#2\else\csname #1#2\endcsname\fi}
\def\fig#1{\ifundefined{fig#1}$\spadesuit$#1\else\csname fig#1\endcsname\fi}
\let\endarg=\par
\def\finish{\def\endarg{\par\endgroup}}
\def\start{\endarg\begingroup}

 \def\beginFROM{\start\parskip=0pt\vskip\baselineskip
\def\finish{\def\endarg{\egroup\par\endgroup}}
  \vbox\bgroup\obeylines\eightpoint\em\finish}

\def\ABSTRACT#1\par{
\vskip 1in {\noindent\sectionsize\sectiontype Abstract.} #1 \par}

\def\TODAY{\number\day~\ifcase\month\or January \or February \or March \or
April \or May \or June
\or July \or August \or September \or October \or November \or December \fi
\number\year\timecount=\number\time
\divide\timecount by 60
}
\newcount\timecount
\def\DRAFT{\def\lmargin##1{\strut\vadjust{\kern-\strutdepth
\vtop to \strutdepth{
\baselineskip\strutdepth\vss\rlap{\kern-1.2 truecm\eightpoint{##1}}}}}
\font\footfont=cmti7
\footline={{\footfont \hfil File:\jobname, \TODAY,  \number\timecount h}}
}
\newbox\strutboxJPE
\setbox\strutboxJPE=\hbox{\strut}
\def\subitem#1#2\par{\vskip\baselineskip\vskip-\ht\strutboxJPE{\item{#1}#2}}
\gdef\strutdepth{\dp\strutbox}
\def\lmargin#1{}
\def\hexnumber#1{\ifcase#1 0\or1\or2\or3\or4\or5\or6\or7\or8\or9\or
 A\or B\or C\or D\or E\or F\fi}
\textfont\msbfam=\tenmsb
\scriptfont\msbfam=\sevenmsb
\scriptscriptfont\msbfam=\fivemsb
\mathchardef\varkappa="0\hexnumber\msbfam7B%
\textfont\msafam=\tenmsa
\scriptfont\msafam=\sevenmsa
\scriptscriptfont\msafam=\fivemsa
\newcount\FIGUREcount \FIGUREcount=0
\newdimen\figcenter
\def\definefigure#1{\global\advance\FIGUREcount by 1%
\NEWDEF{fig}{#1}{Fig.\ \number\FIGUREcount}
\immediate\write16{  FIG \number\FIGUREcount : #1}}
\def\figure#1#2#3#4\cr{\null%
\definefigure{#1}
{\goodbreak\figcenter=\hsize\relax
\advance\figcenter by -#3truecm
\divide\figcenter by 2
\midinsert\vskip #2truecm\noindent\hskip\figcenter
\includegraphics{#1}\vskip 0.8truecm\noindent \vbox{\eightpoint\noindent
{\bf\fig{#1}}: #4}\endinsert}}
\def\figurewithtex#1#2#3#4#5\cr{\null%
\definefigure{#1}
{\goodbreak\figcenter=\hsize\relax
\advance\figcenter by -#4truecm
\divide\figcenter by 2
\midinsert\vskip #3truecm\noindent\hskip\figcenter
\includegraphics{#1}{\hskip\texpscorrection\input #2 }\vskip 0.8truecm\noindent \vbox{\eightpoint\noindent
{\bf\fig{#1}}: #5}\endinsert}}
\def\figurewithtexplus#1#2#3#4#5#6\cr{\null%
\definefigure{#1}
{\goodbreak\figcenter=\hsize\relax
\advance\figcenter by -#4truecm
\divide\figcenter by 2
\midinsert\vskip #3truecm\noindent\hskip\figcenter
\includegraphics{#1}{\hskip\texpscorrection\input #2 }\vskip #5truecm\noindent \vbox{\eightpoint\noindent
{\bf\fig{#1}}: #6}\endinsert}}
\catcode`@=11
\def\footnote#1{\let\@sf\empty 
  \ifhmode\edef\@sf{\spacefactor\the\spacefactor}\/\fi
  #1\@sf\vfootnote{#1}}
\def\vfootnote#1{\insert\footins\bgroup\eightpoint
  \interlinepenalty\interfootnotelinepenalty
  \splittopskip\ht\strutbox 
  \splitmaxdepth\dp\strutbox \floatingpenalty\@MM
  \leftskip\z@skip \rightskip\z@skip \spaceskip\z@skip \xspaceskip\z@skip
  \textindent{#1}\footstrut\futurelet\next\fo@t}
\def\fo@t{\ifcat\bgroup\noexpand\next \let\next\f@@t
  \else\let\next\f@t\fi \next}
\def\f@@t{\bgroup\aftergroup\@foot\let\next}
\def\f@t#1{#1\@foot}
\def\@foot{\strut\egroup}
\def\footstrut{\vbox to\splittopskip{}}
\skip\footins=\bigskipamount 
\count\footins=1000 
\dimen\footins=8in 
\catcode`@=12 

\def\AA{{\script A}}

\def\CC{{\script C}}

\def\LL{{\script L}}
\def\MM{{\script M}}

\def\OO{{\script O}}

\def\HALF{{\textstyle{1\over 2}}}

\def\QEDD{\hfill\smallskip
         \line{$\hfill{\vcenter{\vbox{\hrule height 0.2pt
	\hbox{\vrule width 0.2pt height 1.3ex \kern 1.3ex
		\vrule width 0.2pt}
	\hrule height 0.2pt}}}$
               \ \ \ \ \ \ }
         \bigskip}
\def\QED{$\hfill{\vcenter{\vbox{\hrule height 0.2pt
	\hbox{\vrule width 0.2pt height 1.3ex \kern 1.3ex
		\vrule width 0.2pt}
	\hrule height 0.2pt}}}$\bigskip}
\def\real{{\bf R}}

\def\complex{{\bf C}}

\def\PROOF{\medskip\noindent{\bf Proof.\ }}
\def\REMARK{\medskip\noindent{\bf Remark.\ }}
\def\LIKEREMARK#1{\medskip\noindent{\bf #1.\ }}
\normalbaselineskip=5.25mm
\baselineskip=5.25mm
\parskip=10pt
\beforesectionskipamount=24pt plus8pt minus8pt
\sectionskipamount=3pt plus1pt minus1pt
\def\em{\it}
\tenpoint
\null
\catcode`\@=11
\ifx\amstexloaded@\relax\catcode`\@=\active
 \fi
\catcode`\@=\active
\def\period{\unskip.\spacefactor3000 { }}
%
%
\newbox\noboxJPE
\newbox\byboxJPE
\newbox\paperboxJPE
\newbox\yrboxJPE
\newbox\jourboxJPE
\newbox\pagesboxJPE
\newbox\volboxJPE
\newbox\preprintboxJPE
\newbox\toappearboxJPE
\newbox\bookboxJPE
\newbox\bybookboxJPE
\newbox\publisherboxJPE
\newbox\inprintboxJPE
\def\refclearJPE{
   \setbox\noboxJPE=\null             \gdef\isnoJPE{F}
   \setbox\byboxJPE=\null             \gdef\isbyJPE{F}
   \setbox\paperboxJPE=\null          \gdef\ispaperJPE{F}
   \setbox\yrboxJPE=\null             \gdef\isyrJPE{F}
   \setbox\jourboxJPE=\null           \gdef\isjourJPE{F}
   \setbox\pagesboxJPE=\null          \gdef\ispagesJPE{F}
   \setbox\volboxJPE=\null            \gdef\isvolJPE{F}
   \setbox\preprintboxJPE=\null       \gdef\ispreprintJPE{F}
   \setbox\toappearboxJPE=\null       \gdef\istoappearJPE{F}
   \setbox\inprintboxJPE=\null        \gdef\isinprintJPE{F}
   \setbox\bookboxJPE=\null           \gdef\isbookJPE{F}  \gdef\isinbookJPE{F}
     
   \setbox\bybookboxJPE=\null         \gdef\isbybookJPE{F}
   \setbox\publisherboxJPE=\null      \gdef\ispublisherJPE{F}
}

\def\ref{\refclearJPE}
\def\no#1{\gdef\isnoJPE{T}\setbox\noboxJPE=\hbox{#1}}
\def\by#1{\gdef\isbyJPE{T}\setbox\byboxJPE=\hbox{#1}}
\def\paper#1{\gdef\ispaperJPE{T}\setbox\paperboxJPE=\hbox{#1}}
\def\yr#1{\gdef\isyrJPE{T}\setbox\yrboxJPE=\hbox{#1}}
\def\jour#1{\gdef\isjourJPE{T}\setbox\jourboxJPE=\hbox{#1}}
\def\pages#1{\gdef\ispagesJPE{T}\setbox\pagesboxJPE=\hbox{#1}}
\def\vol#1{\gdef\isvolJPE{T}\setbox\volboxJPE=\hbox{\bf #1}}
\def\preprint#1{\gdef
\ispreprintJPE{T}\setbox\preprintboxJPE=\hbox{#1}}

\def\book#1{\gdef\isbookJPE{T}\setbox\bookboxJPE=\hbox{\em #1}}
\def\publisher#1{\gdef
\ispublisherJPE{T}\setbox\publisherboxJPE=\hbox{#1}}
\def\inbook#1{\gdef\isinbookJPE{T}\setbox\bookboxJPE=\hbox{\em #1}}
\def\bybook#1{\gdef\isbybookJPE{T}\setbox\bybookboxJPE=\hbox{#1}}
\newdimen\refindent
\refindent=5em
\def\endref{\sfcode`.=1000
 \if T\isnoJPE
\hangindent\refindent\hangafter=1
      \noindent\hbox to\refindent{[\unhbox\noboxJPE\unskip]\hss}\ignorespaces
     \else  \noindent    \fi
 \if T\isbyJPE    \unhbox\byboxJPE\unskip: \fi
 \if T\ispaperJPE \unhbox\paperboxJPE\unskip\period \fi
 \if T\isbookJPE {\it\unhbox\bookboxJPE\unskip}\if T\ispublisherJPE, \else.
\fi\fi
 \if T\isinbookJPE In {\it\unhbox\bookboxJPE\unskip}\if T\isbybookJPE,
\else\period \fi\fi
 \if T\isbybookJPE  (\unhbox\bybookboxJPE\unskip)\period \fi
 \if T\ispublisherJPE \unhbox\publisherboxJPE\unskip \if T\isjourJPE, \else\if
T\isyrJPE \  \else\period \fi\fi\fi
 \if T\istoappearJPE (To appear)\period \fi
 \if T\ispreprintJPE Pre\-print\period \fi
 \if T\isjourJPE    \unhbox\jourboxJPE\unskip\ \fi
 \if T\isvolJPE     \unhbox\volboxJPE\unskip\if T\ispagesJPE, \else\ \fi\fi
 \if T\ispagesJPE   \unhbox\pagesboxJPE\unskip\  \fi
 \if T\isyrJPE      (\unhbox\yrboxJPE\unskip)\period \fi
 \if T\isinprintJPE (in print)\period \fi
\filbreak
}

\expandafter\xdef\csname
sIntro1\endcsname{1}
\expandafter\xdef\csname
ethe\endcsname{(1.1)}
\expandafter\xdef\csname
sIntro\endcsname{2}
\expandafter\xdef\csname
eLambda\endcsname{(2.1)}
\expandafter\xdef\csname
eno\endcsname{(2.2)}
\expandafter\xdef\csname
eno2\endcsname{(2.3)}
\expandafter\xdef\csname
elambdamin\endcsname{(2.4)}
\expandafter\xdef\csname
eyes\endcsname{(2.5)}
\expandafter\xdef\csname
sabc\endcsname{3}
\expandafter\xdef\csname
erhoess\endcsname{(3.1)}
\expandafter\xdef\csname
crhoess\endcsname{Definition\ 3.1}
\expandafter\xdef\csname
esigmaess\endcsname{(3.2)}
\expandafter\xdef\csname
esigmapess\endcsname{(3.3)}
\expandafter\xdef\csname
csigmaess\endcsname{Definition\ 3.2}
\expandafter\xdef\csname
econj1\endcsname{(3.4)}
\expandafter\xdef\csname
cconj1\endcsname{Theorem\ 3.3}
\expandafter\xdef\csname
erhoess2\endcsname{(3.5)}
\expandafter\xdef\csname
cpetit\endcsname{Lemma\ 3.4}
\expandafter\xdef\csname
estep1\endcsname{(3.6)}
\expandafter\xdef\csname
sBaladi1\endcsname{4}
\expandafter\xdef\csname
figfigsbaladi1.ps\endcsname{Fig.\ 1}
\expandafter\xdef\csname
figfigsbaladi2.ps\endcsname{Fig.\ 2}
\expandafter\xdef\csname
ehfunct\endcsname{(4.1)}
\expandafter\xdef\csname
exyz\endcsname{(4.2)}
\expandafter\xdef\csname
figfigsbaladi5.ps\endcsname{Fig.\ 3}
\expandafter\xdef\csname
figfigsbaladi4.ps\endcsname{Fig.\ 4}
\expandafter\xdef\csname
sBaladiskew\endcsname{5}
\expandafter\xdef\csname
figfigsbaladi3.ps\endcsname{Fig.\ 5}
\expandafter\xdef\csname
sBaker\endcsname{6}
\expandafter\xdef\csname
figfigsbaker1.ps\endcsname{Fig.\ 6}
\expandafter\xdef\csname
eu\endcsname{(6.1)}
\expandafter\xdef\csname
eip1\endcsname{(6.2)}
\expandafter\xdef\csname
figfigsbaker2.ps\endcsname{Fig.\ 7}
\expandafter\xdef\csname
eCI000\endcsname{(6.3)}
\expandafter\xdef\csname
svv\endcsname{7}
\expandafter\xdef\csname
econtraintes\endcsname{(7.1)}
\expandafter\xdef\csname
edet\endcsname{(7.2)}
\expandafter\xdef\csname
cbbb\endcsname{Proposition\ 7.1}
\expandafter\xdef\csname
sCI\endcsname{8}
\expandafter\xdef\csname
eci999\endcsname{(8.1)}
\expandafter\xdef\csname
eci0\endcsname{(8.2)}
\expandafter\xdef\csname
eci2\endcsname{(8.3)}
\expandafter\xdef\csname
cCI\endcsname{Theorem\ 8.1}
\expandafter\xdef\csname
eci2a\endcsname{(8.4)}
\expandafter\xdef\csname
cCII\endcsname{Theorem\ 8.2}
\expandafter\xdef\csname
cCIIII\endcsname{Conjecture\ 8.3}
\expandafter\xdef\csname
enewer\endcsname{(8.5)}
\expandafter\xdef\csname
eunequation\endcsname{(8.6)}
\expandafter\xdef\csname
epressure\endcsname{(8.7)}
\expandafter\xdef\csname
eci21\endcsname{(8.8)}
\expandafter\xdef\csname
econjugate\endcsname{(8.9)}
\expandafter\xdef\csname
evariance\endcsname{(8.10)}
\expandafter\xdef\csname
cconjugation\endcsname{Theorem\ 8.4}
\expandafter\xdef\csname
cx\endcsname{Corollary\ 8.5}
\expandafter\xdef\csname
etemporary\endcsname{(8.11)}
\expandafter\xdef\csname
sGL\endcsname{9}
\expandafter\xdef\csname
eGL\endcsname{(9.1)}
\expandafter\xdef\csname
cGL\endcsname{Theorem\ 9.1}
\expandafter\xdef\csname
eruelle\endcsname{(9.2)}
\expandafter\xdef\csname
eruelle2\endcsname{(9.3)}
\expandafter\xdef\csname
cCGL\endcsname{Corollary\ 9.2}
\expandafter\xdef\csname
cdifficult\endcsname{Question\ 9.3}
\expandafter\xdef\csname
sGC\endcsname{10}
\expandafter\xdef\csname
ecorrelation\endcsname{(10.1)}
\expandafter\xdef\csname
ehyp1\endcsname{(10.2)}
\expandafter\xdef\csname
erhodef\endcsname{(10.3)}
\expandafter\xdef\csname
chyp\endcsname{Assumption\ 10.1}
\expandafter\xdef\csname
figfigswww.eps\endcsname{Fig.\ 8}
\expandafter\xdef\csname
evarphi\endcsname{(10.4)}
\expandafter\xdef\csname
ea1\endcsname{(10.5)}
\expandafter\xdef\csname
ea2\endcsname{(10.6)}
\expandafter\xdef\csname
exyzz\endcsname{(10.7)}
\expandafter\xdef\csname
cass3\endcsname{Assumption\ 10.2}
\expandafter\xdef\csname
cC1\endcsname{Conjecture\ 10.3}
\expandafter\xdef\csname
cremark\endcsname{Question\ 10.4}
\expandafter\xdef\csname
eg1def\endcsname{(10.8)}
\expandafter\xdef\csname
eKoop\endcsname{(10.9)}
\expandafter\xdef\csname
eintpart\endcsname{(10.10)}
\expandafter\xdef\csname
egcond\endcsname{(10.11)}
\expandafter\xdef\csname
egcond2\endcsname{(10.12)}
\expandafter\xdef\csname
eb1\endcsname{(10.13)}
\expandafter\xdef\csname
eabc\endcsname{(10.14)}
\expandafter\xdef\csname
ex1\endcsname{(10.15)}
\expandafter\xdef\csname
ex2\endcsname{(10.16)}
\expandafter\xdef\csname
ealmost\endcsname{(10.17)}
\expandafter\xdef\csname
ebunch1\endcsname{(10.18)}
\expandafter\xdef\csname
ebunch2\endcsname{(10.19)}
\expandafter\xdef\csname
csmooth\endcsname{Theorem\ 10.5}

\normalbaselineskip=12pt
\baselineskip=12pt
\parskip=0pt
\parindent=22.222pt
\beforesectionskipamount=24pt plus0pt minus6pt
\sectionskipamount=7pt plus3pt minus0pt
\overfullrule=0pt
\hfuzz=2pt
\nopagenumbers
\headline={\ifnum\pageno>1 {\hss\tenrm-\ \folio\ -\hss} \else
{\hfill}\fi}
\font\titlefont=ptmb at 14 pt

\font\toplinefont=cmcsc10
\font\pagenumberfont=ptmb at 10pt

\newdimen\itemindent\itemindent=1.5em

\def\textindent#1{\indent\llap{#1\enspace}\ignorespaces}
\def\item{\par\noindent
\hangindent\itemindent\hangafter=1\relax
\setitemmark}
\def\setitemindent#1{\setbox0=\hbox{\ignorespaces#1\unskip\enspace}%
\itemindent=\wd0\relax
\message{|\string\setitemindent: Mark width modified to hold
         |`\string#1' plus an \string\enspace\space gap. }%
}
\def\setitemmark#1{\checkitemmark{#1}%
\hbox to\itemindent{\hss#1\enspace}\ignorespaces}
\def\checkitemmark#1{\setbox0=\hbox{\enspace#1}%
\ifdim\wd0>\itemindent
   \message{|\string\item: Your mark `\string#1' is too wide. }%
\fi}
\newcount\FOOTcount \FOOTcount=0
\def\myfoot#1{\global\advance\FOOTcount by 1\footnote{${}^{\number\FOOTcount}$}{#1}}


\newread\refs
\def\ref#1{#1}
\def\PC99{{\tt *****PC99:}}


\expandafter\xdef\csname refBaladi:1989\endcsname{1}\relax
\expandafter\xdef\csname refBaladi2000\endcsname{2}\relax
\expandafter\xdef\csname refBowen1975\endcsname{3}\relax
\expandafter\xdef\csname refColletIsola1991\endcsname{4}\relax
\expandafter\xdef\csname refEE\endcsname{5}\relax
\expandafter\xdef\csname refGundlachLatushkin:2001\endcsname{6}\relax
\expandafter\xdef\csname refHPS1977\endcsname{7}\relax
\expandafter\xdef\csname refKato1984\endcsname{8}\relax
\expandafter\xdef\csname refKatokHasselblatt1995\endcsname{9}\relax
\expandafter\xdef\csname refKeller1979\endcsname{10}\relax
\expandafter\xdef\csname refKeller1984\endcsname{11}\relax
\expandafter\xdef\csname refMattila1995\endcsname{12}\relax
\expandafter\xdef\csname refRousseau-Egele1983\endcsname{13}\relax
\expandafter\xdef\csname refRuelle1978\endcsname{14}\relax
\expandafter\xdef\csname refRuelle1986PRL\endcsname{15}\relax
\expandafter\xdef\csname refRuelle1987\endcsname{16}\relax
\expandafter\xdef\csname refYoung1999\endcsname{17}\relax

\def\cite#1{\csname ref#1\endcsname}
\def\bibitem#1{\item{[\csname ref#1\endcsname]}}
\def\newblock{}
\def\endref{}
\def\begin#1#2{}
\let\ref=\cite
\def\supapprox{\mathchar "0\hexnumber\msafam26    }
\let\epsilon=\varepsilon
\let\rho=\varrho
\newcount\FOOTcount \FOOTcount=0
\def\myfoot#1{\global\advance\FOOTcount by 1\footnote{${}^{\number\FOOTcount}$}{#1}}
\let\truett=\tt
\fontdimen3\tentt=2pt\fontdimen4\tentt=2pt
\def\tt{\hfill\break\null\kern -2truecm\truett ************ }
\def\CC{{\cal C}}
\def\text#1{\leavevmode\hbox{#1}}
\headline
{\ifnum\pageno>1 {\toplinefont Liapunov Multipliers and Decay of Correlations}
\hfill{\pagenumberfont\folio}\fi}
\def\ie{{\it i.e.}}
\def\MM{{\cal M}}
\def\WW{{\cal W}}
\def\refb#1{[\ref{#1}]}
\def\Oun{{\cal O}(1)}
\def\map{f}
\def\eg{{\it e.g.},}
\def\var{{\rm var }}
\def\h{h}
\def\rhoess{\rho_{\rm ess}}
\def\lambdastar{\lambda_{\rm ess}}
\def\sigmapess{\sigma_{\rm p-ess}}
\def\sigmaess{\sigma_{\rm ess}}
\def\sigmasp{\sigma_{\rm sp}}
\def\EDR{essential decorrelation radius}
\def\EDRA{essential decorrelation rate}
\def\ESR{essential spectral radius}
\def\PESR{point-essential spectral radius}
\def\V{V}
\def\T{{\cal T}}
\def\S{{\cal S}}
\noindent
{\titlefont{\centerline{Liapunov Multipliers }}}
\vskip 0.5truecm
{\titlefont{\centerline{{and Decay of Correlations in Dynamical Systems}}}}
\vskip 0.5truecm
{\it{\centerline{ P. Collet${}^{1}$ and J.-P. Eckmann${}^{2,3}$}}}
\vskip 0.3truecm
{\eightpoint
\centerline{${}^1$Centre de Physique Th\'eorique, Laboratoire CNRS UMR
7644,
Ecole Polytechnique, F-91128 Palaiseau Cedex, France}
\centerline{${}^2$D\'ept.~de Physique Th\'eorique, Universit\'e de Gen\`eve,
CH-1211 Gen\`eve 4, Switzerland}
\centerline{${}^3$Section de Math\'ematiques, Universit\'e de Gen\`eve,
CH-1211 Gen\`eve 4, Switzerland}
}
\vskip 0.5truecm
\centerline{\it Dedicated to Gianni Jona-Lasinio in gratitude for his
many encouragements.}
\vskip 0.5truecm
{\eightpoint\narrower\baselineskip 11pt
\LIKEREMARK{Abstract}The essential decorrelation rate of a hyperbolic
dynamical system is 
the decay rate of time-correlations one expects to see stably for
typical observables 
once resonances are projected out. We define and illustrate these notions
and study the conjecture that for
observables in $\CC^1$, the
essential decorrelation rate is never faster than what is dictated by the {\em
smallest} unstable Liapunov multiplier.
}
\SECT{Intro1}{Introduction}The purpose of this paper is a discussion
of the relation between the decay of time-correlations and the Liapunov
exponents of dynamical systems. It is well-known that if a system has
vanishing Liapunov exponents, in general the decay of correlations can
be arbitrarily slow. Here, we study the case when the Liapunov
exponents are all {\em different} from 0.
The decay of time-correlations in a
dynamical system depends in general on the type of observable one
considers. We will explain below why, in our view, the class $\CC^1$
of once differentiable observables is a natural and useful
choice. Whatever the choice, it implies a notion of {\em essential
decorrelation rate}, also to be defined below. Its intuitive meaning is
perhaps best understood in terms of {\em resonances} or {\em
improvable decorrelation rates} [\ref{Ruelle1986PRL},\ref{Ruelle1987}]. 
These are rates which will be seen
usually for any randomly picked observable and which are {\em slower}
than the essential ones. But they are improvable in the following
sense. For any $\epsilon >0$,
there is a finite dimensional subspace of such observables, and if we take any
other observable in the complement of this subspace, we will see the
\EDRA{} within $\epsilon $. It is precisely called essential, because no
further {\em finite dimensional} restriction of observables will lead to a
faster decorrelation rate. 

We will first define with mathematical precision an {\em \EDR} $\rhoess$ which
is the inverse of the essential decorrelation rate $\lambdastar$.
We will then show by means of some
examples that systems with {\em improvable decorrelation rates really
exist}. We
then address 
the question of the \EDR. We will study for
observables in $\CC^1$ and for several expanding systems
the validity of the inequality
$$
\rhoess \,\equiv\, 1/\lambdastar \,\ge\, 1/\lambda_{\min}~,
\EQ{the}
$$
where $\log\lambda_{\min}$ is the {\bf smallest} positive Liapunov
multiplier.\myfoot{The Liapunov exponent is the logarithm of the
Liapunov multiplier}
We also argue that in many cases the inequality above is strict, so
that the (essential) decay rate of correlations is even {\em slower}
than what is suggested by the smallest positive Liapunov
multiplier.\myfoot{It is somewhat anti-intuitive that the lowest and
not the largest Liapunov exponent matters, when
compared with the idea that
Liapunov exponents are separation rates, but the reader
should note that decay rates are really infinite time quantities, and
the fast local separation of orbits only works for a short time, and
only for a few avoidable observables.}

Our paper deals thus with lower bounds not only on the essential
spectrum, but also on the \EDR. For related work, see [\ref{Young1999}].

\SECT{Intro}{Setup}We consider throughout a smooth
manifold $\MM$ of dimension $d$, 
and a (piecewise) smooth map $f$ of $\MM$
into itself. The differential  of
$f$ (a $d\times d$ matrix)
is denoted $Df$, and $Df(x)$ or $\left . Df\right |_x$ when evaluated
at the point $x$. Two 
quantities of interest in ``chaotic'' systems are the {\em Liapunov
multipliers} and the correlation functions. The Liapunov multipliers are 
obtained by considering first the matrices
$$
\Lambda_n(x)\,=\, Df\bigl(f^{n-1}(x)\bigr)\cdot
Df\bigl(f^{n-2}(x)\bigr)\cdots
Df\bigl(x\bigr)\,\equiv \prod_{i=0}^{n-1} Df\bigl(f^i(x)\bigr)~.
$$
By Oseledec' theorem, given an invariant measure $\nu$, 
the Liapunov multipliers are then the eigenvalues of
the matrix 
$$
\lim_{n\to\infty }\bigl(\Lambda_n (x)^*\Lambda_n (x)\bigr)^{1\over 2n}~,
\EQ{Lambda}
$$
(which exists
$\nu$ almost everywhere)\myfoot{We always write $\lambda $ for the Liapunov
multiplier, 
and $\log \lambda $ for the corresponding Liapunov exponent.}. 
If the system is in addition ergodic with respect to the
invariant measure,
then these eigenvalues are $\nu$-almost surely
independent of $x$. Note that the Liapunov multipliers will in general
depend on $\nu$ when there are several invariant measures. We will
call these multipliers\myfoot{They depend on the invariant measure.}
$$
\lambda_1\,\ge\,\lambda_2\,\ge\,\cdots\,\ge\,\lambda_d~.
$$

Recall also that for SRB measures the limit in \equ{Lambda} exists Lebesgue almost surely
(in the basin of the measure) and not only on the support of the
measure $\nu$
which may be of Lebesgue measure zero.

A second quantity of interest are correlation functions.
Consider two observables, $F$ and $G$, which are functions on $\MM$
taking real values. 
Here, and throughout the paper, we will assume that $F$ and $G$ have
zero mean.
Then we can form the correlation functions
$$
S_k(x)\,=\,\lim_{n\to\infty}{1\over n}\sum_{j=0}^{n-1}
F\bigl(f^{j+k}(x)\bigr)G\bigl(f^{j} (x)\bigr)~.
$$
Again, 
$S_k(x)$ is Lebesgue almost surely independent of $x$ and is also equal
to
$$
S_k\,=\,\int d\mu (x) F\bigl(f^k(x)\bigr)\,G(x) ~,
$$
where $\mu$ is the SRB measure (assuming it exists).

A question of interest is the relation between the rate of decay
of $S_k$ as $k\to\infty $ and the Liapunov multipliers. A tempting
idea is to argue that since the orbits seem to separate at a rate
$\lambda_1$ (per unit time) the observables should decorrelate like
$$
|S_k|\,\supapprox\, {C\over \lambda_1^k}~,
\EQ{no}
$$
for some constant $C$. While this property is in a way true for {\em
short times} (small $k$) because generally there is a component of the
observables which ``feels'' the fast rate of the largest Liapunov
multiplier,  the purpose of this paper is to show that 
\equ{no} {\bf does not hold} asymptotically 
in general.

First of all, closer scrutiny of the separation argument given above
indicates that the expected behavior of $S_k$ should be dictated not by the
largest Liapunov multiplier, but rather by the {\em smallest above} 1:
$$
|S_k|\,\supapprox\, {C\over \lambda_{\min}^k}~,
\EQ{no2}
$$
where\myfoot{We do not consider systems with Liapunov multipliers
equal to 1, where it is known that the decorrelation rate may not even
be exponential.} 
$$
\lambda_{\min}\,=\,\min\{ \lambda_i~:~ \lambda_i>1\}~.
\EQ{lambdamin}
$$
We will see that \equ{no2} holds for certain special examples, but for
a general map 
the Equation
\equ{no2} {\bf cannot be an equality} for generic observables in
$\CC^1$, even if we 
avoid the resonances. 
Namely, we expect for maps $f$ with non-constant derivative and for
observables in $\CC^1$ an inequality
$$
|S_k|\,\supapprox\, {C\over \lambdastar ^k}~,
\EQ{yes} 
$$
with $1<\lambdastar <\lambda_{\min}$: {\bf In general, the
decorrelation is
slower than $C/\lambda_{\min}^k$}. Furthermore, $\lambdastar $ is a much
stronger barrier to decay than the resonances: Only a very radical
restriction of the observables (to a subspace of $\CC^1$ with {\em
infinite} codimension) will in general lead to a faster decay.

The purpose of this paper is to clarify the issues related to these questions.

\SECT{abc}{The Essential Decorrelation Radius}In this section we
define the \EDR{} $\rhoess$.
The essential decorrelation rate $\lambdastar$ is then defined by
$$
\lambdastar\,=\,1/\rhoess~,
$$
so that the correlation functions $S_k$ will basically decay like
$\lambdastar^{-k}=\rhoess^k$.

The definition of $\rhoess$ depends on two Banach\myfoot{Hilbert spaces
are not adequate since we work with functions in $\CC^1$.}
spaces
 $X$ and $Y$,
with $X$ a subspace of the dual of $Y$,
The reader should think of $X$ and $Y$ as the Banach space of $\CC^1$
functions with the norm $\|h\|=\sup_x |h(x)| +\sup_x |Dh(x)|$, but we
will need more complicated spaces later.
We denote by $\langle\;,\;\rangle$ the 
continuous bilinear form on $Y\times X$ which is the restriction of
the pairing of $Y$ with its dual.
\CLAIM{Definition}{rhoess}{Let $U$ be a bounded linear
operator on $X$. 
We define the \EDR{} of\/ $U$ on $X,Y$ by
$$
\rhoess(X,Y,U)
\,=\,\inf_{{\rm Codim } M<\infty\,\atop
{\rm Codim }\, M'<\infty}\limsup_{n\to\infty}\left(\sup_{x\in M\backslash\{0\}
,\,\, y\in M'\backslash\{0\}}
{|\langle y,U^{n}x\rangle| \over \|x\|\|y\|}\right)^{1/n}\;.
\EQ{rhoess}
$$
}

\REMARK{The reason we want the space $X$ to be invariant under $U$ is to make
connection later on with the spectral radius. This will force us
to use spaces $X$ whose definitions are a little involved. Although such
a problem does not seem to appear in the definition of the correlation
function, it is hidden in the duality relation between the two
observables.}

The idea of \clm{rhoess} is to peal-out the
various finite dimensional spectral subspaces corresponding to
eigenvalues
outside of the essential
spectral radius.

The essential spectral radius $\sigmaess$ of $U$ on $X$ can be defined in many
equivalent ways, see \eg ~[\ref{EE}, p.~44]. For our purpose the following one
will be used ($r_{\rm e2}(U)$ in [\ref{EE}]):
\CLAIM{Definition}{sigmaess}{Let $U$ be a continuous linear
operator on $X$. We define the \ESR{} by 
$$
\sigmaess(X,U)
\,=\,\sup \{ |\lambda |~:~ {\rm dim}~{\rm Ker} (U-\lambda {\bf
1})=\infty \text{ or ~} (U-\lambda {\bf
1})X~\text{ is not closed }\}~,
\EQ{sigmaess}
$$
and the \PESR{} by 
$$
\eqalign{
\sigmapess(X,U)
\,=\,\sup \{ |\lambda |~:~& \lambda \in\complex{}\text{ is an
accumulation point of eigenvalues}\cr
&\text{or an eigenvalue of infinite multiplicity} \}~.\cr}
\EQ{sigmapess}
$$
}

\CLAIM{Theorem}{conj1}{Let $U$ be a continuous linear
operator on $X$. If $X\subset Y^*$, then
$$
\rhoess(X,Y,U)\,\ge\,\sigmapess(X,U)~.
\EQ{conj1}
$$
}

\REMARK It would be much nicer if we knew that
$\rhoess(X,Y,U)\,\ge\,\sigmaess(X,U)$.
Some of the difficulties of this paper would disappear, and the
considerations of \sec{CI} and \sec{GL} would immediately give the
inequality
\equ{yes}. Nevertheless, \clm{conj1} is still somewhat
useful because information on $\sigmapess$
is relatively easy to get at.
One might be tempted to conjecture that
$\rhoess(X,Y,U)\ge\sigmaess(X,U)$.
However, we found no proof, since
we do not know those $\lambda $ for which $U-\lambda $ has closed
range.
On the other hand, for those $U$ and $X$ we will consider, we shall find
$\sigmapess(X,U)=\sigmaess(X,U)$, so that in the end, we still have
the more useful inequality $\rhoess(X,Y,U)\ge\sigmaess(X,U)$ in those cases.

\LIKEREMARK{Proof of \clm{conj1}}We first use the following
\CLAIM{Lemma}{petit}{With the notations of \clm{rhoess} one has the
identity
$$
\eqalign{
\inf_{{\rm Codim } M<\infty\,\atop
{\rm Codim }\, M'<\infty}&\limsup_{n\to\infty}\left(\sup_{x\in M\backslash\{0\}
,\,\, y\in M'\backslash\{0\}}
{|\langle y,U^{n}x\rangle| \over \|x\|\|y\|}\right)^{1/n}\cr
\,=\,&\inf_{{\rm Codim } M<\infty,~ M~\rm{closed}\atop
{\rm Codim }\, M'<\infty,~ M'~
\rm{closed}}\limsup_{n\to\infty}\left(\sup_{x\in M\backslash\{0\} 
\,,\, y\in M'\backslash\{0\}}
{|\langle y,U^{n}x\rangle| \over \|x\|\|y\|}\right)^{1/n}\;.}
\EQ{rhoess2}
$$
}

\REMARK The proof of this lemma will be given in the Appendix. The
problem of non-closed subspaces is a well-known nuisance in
controlling intersection, see \eg~ [\ref{Kato1984}, footnote 2,
p.~132]. The above lemma helps
avoiding these esoteric problems.

We consider a complex number
$\lambda\neq0$ together with a sequence $(\lambda_{j})$ of complex numbers
converging to $\lambda$ (some terms of the sequence  and possibly
infinitely many  may be equal to $\lambda$), which are
eigenvectors of $U$ in $X$ associated to the sequence of independent
 eigenvectors $(e_{j})$. We claim that $\rhoess
\ge|\lambda|$. To prove this we will show that for any subspaces
$M\subset X$ and $M'\subset Y$, both of finite codimension, we have
$$
|\lambda|\le\limsup_{n\to\infty}\left(
\sup_{ \scriptstyle x\in M,  \scriptstyle y\in M'}
{\big|\langle y,U^{n}x\rangle\big|\over\|y\|\|x\|}\right)^{1/n} \;.
\EQ{step1}
$$ 
By \clm{petit}, it is enough to show this for closed subspaces $M$ and $M'$.
Let $\epsilon>0$ and denote by $s$ and $s'$ the codimensions of $M$ and
$M'$, respectively. Let $W$ be a subspace generated by $s+s'+1$
vectors among the infinite sequence $(e_{j})$ with respective eigenvalues
of modulus larger than $|\lambda|-\epsilon$. From our hypothesis, this
is always possible. We have $\dim(W\cap M)\ge s'+1$, see [\cite{Kato1984},
problem 1.42, p.~142].

From [\ref{Kato1984}, Lemma 1.40, p.~141]
we conclude that (in $Y^*$) one has $s'=\dim {M'}^{\perp}$  and
therefore, since $X\subset Y^*$ by assumption, we find
$$
W\cap M\not\subset {M'}^{\perp}\;.
$$ 
This implies that there are a $w\in W\cap M$ and a $v\in M'$ such that
$\langle v,w\rangle\neq 0$. Therefore, there is at least one
$e_{\ell}$ 
among those
generating $W$ for which $\langle v,e_{\ell}\rangle\neq0$. Since the associated
eigenvalue $\lambda_{\ell}$ satisfies $|\lambda_{\ell}|\ge|\lambda|-\epsilon$, 
we get
$$
\limsup_{n\to\infty}\left(
\sup_{\scriptstyle x\in M\atop\scriptstyle y\in M'}
{\big|\langle y,U^{n}x\rangle\big|\over\|y\|\|x\|}\right)^{1/n}
\ge \limsup_{n\to\infty}\left(
{\big|\langle v,U^{n}e_{\ell}\rangle\big|\over\|v\|\|e_{\ell}\|}\right)^{1/n}=|\lambda_{\ell}|\ge
|\lambda|-\epsilon\;.
$$
We conclude that for any $\epsilon>0$,
$$
\rhoess\ge |\lambda|-\epsilon~.
$$
Since $\epsilon >0$ is arbitrary, we conclude $\rhoess\ge |\lambda |$
as asserted.
\clm{conj1} follows immediately from the definition of $\sigmapess$.

\SECT{Baladi1}{Baladi Map in 1 Dimension}In this section
we focus on {\em resonances}, by giving a 1-dimensional example.
In \sec{Baladiskew} we give a 2-dimensional, area-preserving example
and in \sec{vv} we show how this example can be generalized to unequal
slopes. This provides then an example with resonances {\em and} for
which the decay rate is
{\em not} 
given by $1/\lambda_{\min}$, but by $1/\lambdastar $ as explained in the
Introduction and in \sec{Intro}.

\figurewithtexplus{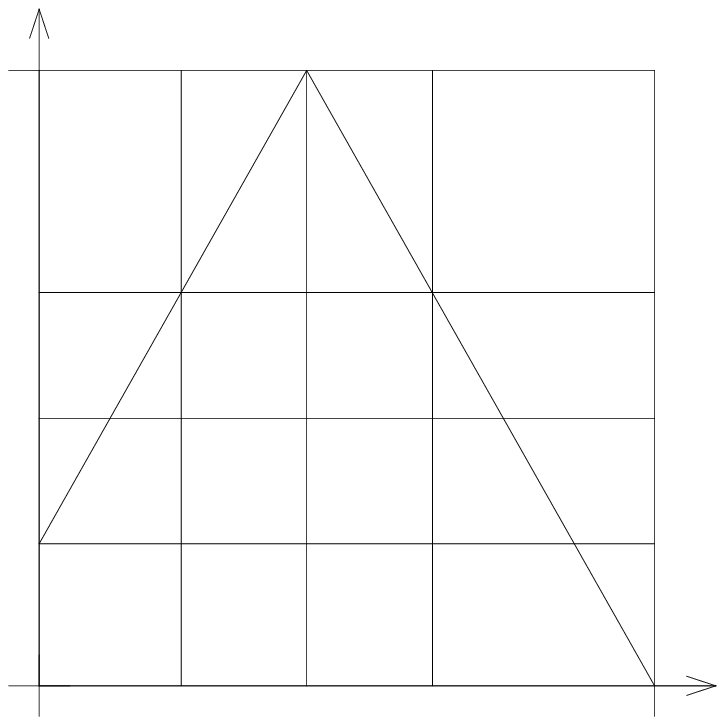}{baladi1.tex}{9}{10}{-0.9}{The
graph of the Baladi map.}\cr
Since there is only one Liapunov multiplier in dimension 1, we shall
write $\lambda $ instead of $\lambda_{\min}$.
There are many maps of the interval
with a slope of constant modulus 
$\lambda >1$, which are Markov and which have resonances in the
correlation function. By a systematic search, Baladi [\ref{Baladi:1989}]
found the simplest
such map, whose partition has only four pieces. \noindent
The map (which we will call $f$) is drawn in 
\fig{figsbaladi1.ps} and is defined by 
$$
f(x)\,=\,\cases{
\hphantom{-}\lambda (x-x_{\rm c})+1 ~,& if $x\le x_{\rm c}~,$\cr
-\lambda (x-x_{\rm c})+1~,& if $x\ge x_{\rm c}~,$\cr
}
$$
where $x_c={2\lambda ^2\over
1+\lambda }+{1-2\lambda ^2\over \lambda ^2}$.
Note that it has a slope $\pm \lambda $,
where $\lambda >1$. Baladi obtained $\lambda $ as follows: 
If we call $P_1,\dots ,P_4$ the four pieces of the partition of
$[0,1]$ as shown on the bottom of \fig{figsbaladi1.ps}, we see
that $f(P_1)=P_2\cup P_3$,  $f(P_2)=f(P_3)=P_4$, and $f(P_4)=P_1\cup
P_2 \cup P_3$.
Therefore, the transition matrix $M$ (the Markov matrix) defined by
$M_{i,j}=1$ if $P_{j}\subset f(P_{i})$ and zero otherwise is given by:
$$
M\,=\,\left (\matrix {0 &0 &0& 1\cr
1&0&0& 1\cr
1&0&0&1\cr
0&1&1&0\cr}\right )~.
$$
Its characteristic polynomial is
$$
\lambda ^4-2\lambda ^2-2\lambda~,
$$
and its eigenvalues are 
$$
\lambda \,\approx\,1.76929~,\quad \lambda_{{\rm r},\pm}\,\approx\,
-0.884846\pm i\, 0.58973~,\text{ and \ }0~.
$$
The reader will check easily that the maximal eigenvalue is the right choice
of $\lambda $.
The correlation functions are given
by
$$
\eqalign{
S_k\,=\,\int dx\, F(x)G\bigl(f^k(x)\bigr) h(x)~,
}
$$
where the density $h$ of the invariant measure (which is unique among the
absolutely continuous invariant measures) is given by
\figurewithtexplus{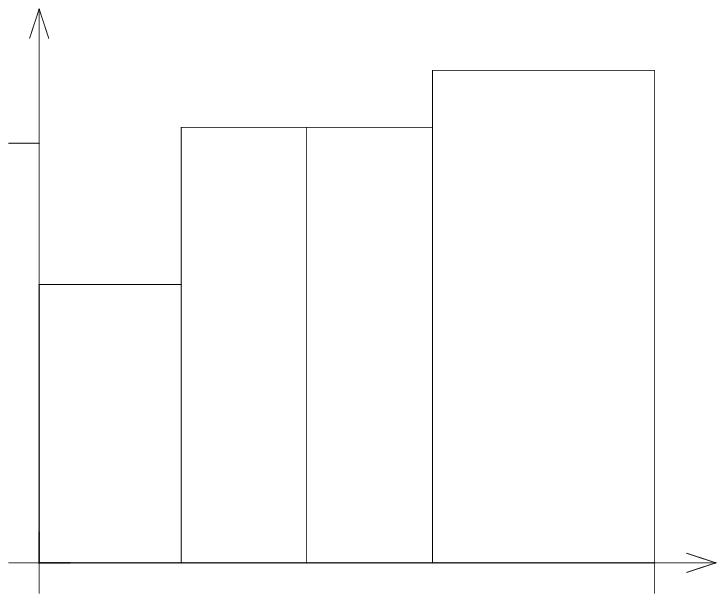}{baladi2.tex}{8}{10}{-0.9}{The
density $h$ of the invariant measure. The normalization factor is
$N=(2\lambda ^3-\lambda -2)/\lambda ^2$.}\cr
$$
h(x)\,=\,
\cases{
\alpha \equiv\lambda ^2/N~,& if $x<2\lambda ^2/(1+\lambda )\equiv x_1~,$\cr
\beta \equiv\lambda (1+\lambda )/N~,& if $2\lambda ^2/(1+\lambda
)<x<2/\lambda ^2\equiv x_2~,$\cr
\gamma \equiv2(1+\lambda )/N~,& if $2/\lambda ^2<x<1~,$\cr
}
\EQ{hfunct}
$$
and $N=(2\lambda ^3-\lambda -2)/\lambda ^2$ is a normalization.
\noindent Changing variables to $y=f^{-1}(x)$ one gets
$$
S_k\,=\,\int dy\, \bigl(P^k (Fh)\bigr)(y) G(y)~,
$$ where $P$ is the Perron-Frobenius operator
$$
\bigl(Pg\bigr)(y)\,=\,\sum_{x:f(x)=y}{g(x)\over |f'(x)|}~.
$$
Note that since $|f'(x)|\equiv\lambda $ for our example,
the Perron-Frobenius operator in this case equals $\lambda ^{-1}M$
when acting on
functions which are constant on the four pieces of the Markov
partition.
Therefore, on that space, its eigenvalues are given by
$$
1~,\quad {\lambda_{{\rm r},\pm}\over \lambda }\,\approx\,
{-0.884846\pm i\, 0.58973\over 1.76929}~,\text{ and \ }0~.
$$
It follows that for generic observables {\em the correlation functions
decay like} 
$$
|S_k|\,\supapprox\,C \left |{\lambda_{{\rm r},\pm} \over \lambda } \right
|^k~.
\EQ{xyz}
$$
This decay rate is {\em slower} than $C |1/\lambda |^k$ because
$|\lambda_{{\rm r},\pm}|\,\approx\,1.06320$. We illustrate these
findings by numerical experiments in \fig{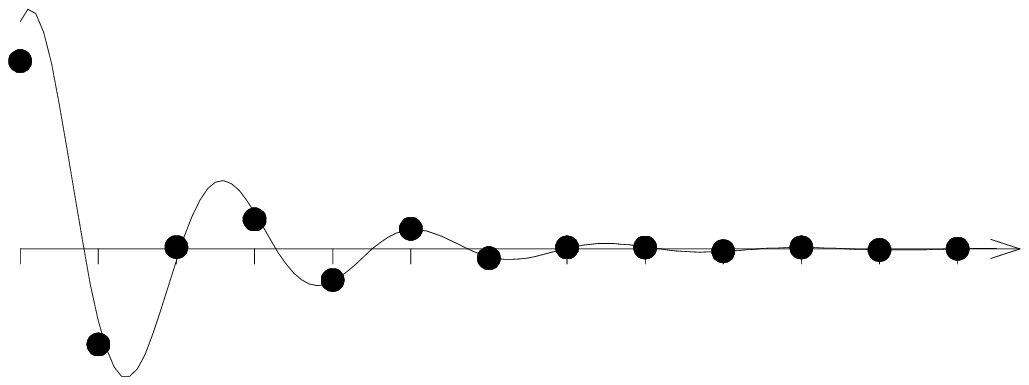} and
\fig{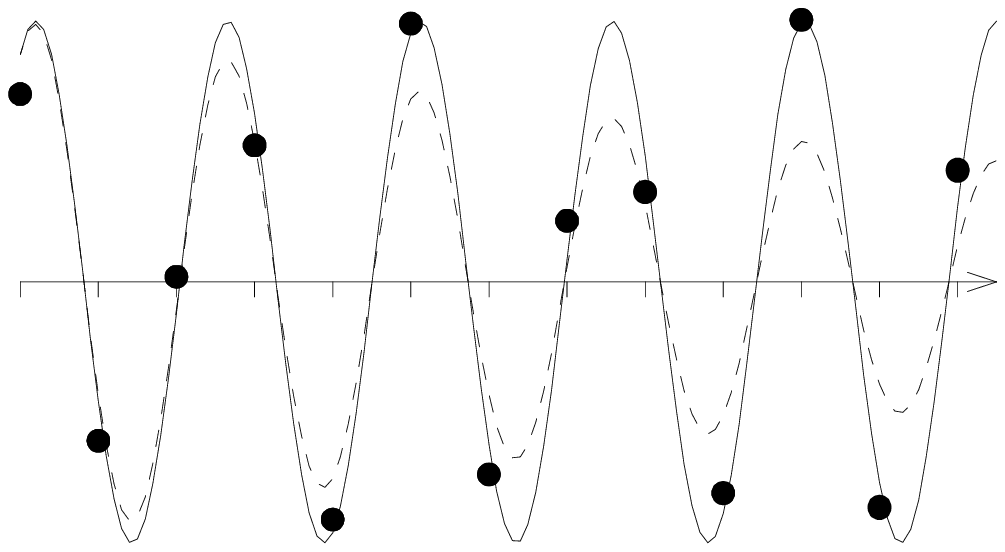}. The question is now whether $C\ne0$. The matrix
$M$ has an eigenvector $v_1=(\alpha ,\beta ,\beta ,\gamma)$ as defined
in \equ{hfunct} corresponding to the eigenvalue $\lambda $, a 2
dimensional eigen-subspace corresponding to the
eigenvalues $\lambda_{{\rm r},\pm}$ (spanned by some vectors $v_2$ and $v_3$),
and a fourth eigendirection $v_4=(0,1,-1,0)$ corresponding to the
eigenvalue 0. These are also eigenspaces for $\lambda ^{-1}P$ . We see
that {\em if the function $F\cdot h$ does not have any component in the
subspace spanned by $v_2$ ,$v_3$, then $C=0$, and the decay of $S_k$
is faster than described in \equ{xyz}}.\myfoot{Strictly speaking, we
have shown this only for functions which are constant on the pieces of
the partition. The proof of the general case is left to the reader.}
In all other cases, $C\ne 0$
and \equ{xyz}  describes the relevant decay rate. We will therefore
say that $\lambda_{{\rm r},\pm}/\lambda $ are {\bf resonances},
see [\ref{Ruelle1986PRL},\ref{Ruelle1987}],
because they can be
avoided by choosing observables (with zero average) 
in a subspace of codimension 2.

\figurewithtexplus{figsbaladi5.ps}{baladi5.tex}{7.5}{10.6}{-3}{A
numerical study of the correlation function $S_k$ for the
1-dimensional Baladi map, from $3\cdot 10^7$ data points. The
continuous graph is the theoretical curve, ${\rm const.~Re}(\lambda
_{{\rm r},+}/\lambda )^k$.}\cr

\figurewithtexplus{figsbaladi4.ps}{baladi4.tex}{8}{10.6}{-2}{The
same data as in \fig{figsbaladi5.ps} but now scaled vertically by
$\bigl|\lambda_{{\rm r},+}/\lambda \bigr|^{-k}$. Superposed is the (dashed)
curve scaled by 
$(1/\lambda) ^{-k}$ which shows clearly the difference between the
decay rate of the resonance $\lambda_{{\rm r},\pm}/\lambda $ and that of the inverse
of the Liapunov multiplier which is $1/\lambda $. }\cr
\SECT{Baladiskew}{A Skew Product Using a Baladi Map}Using the Baladi map $f$ of
the preceding section, we can construct a new map, $\Phi$ which is 
area-preserving, invertible, hyperbolic, and has a resonance (the same
as in \sec{Baladi1}).
The map is defined as in \fig{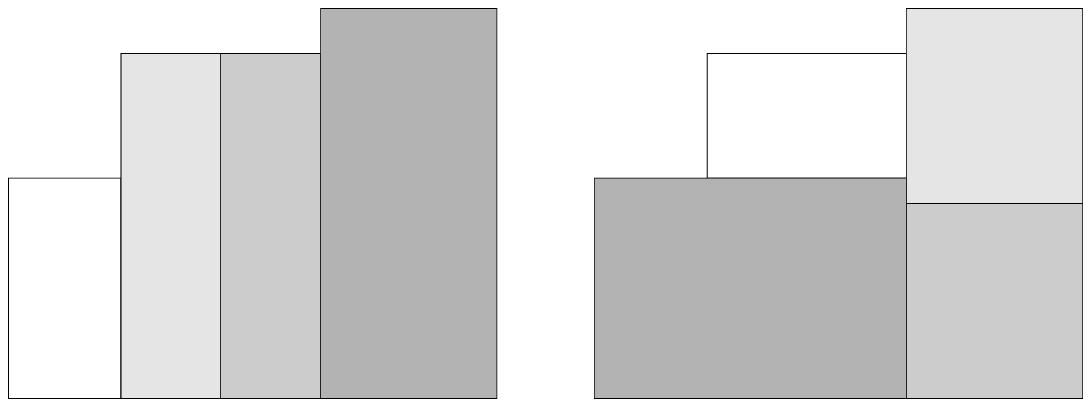}.
In formulas:
$$
\Phi(x,y)\,=\,\bigl(f(x),y/\lambda +t(x)/N\bigr)~,
$$
with $\lambda $ as in the previous section and where $t$ is given by
\figurewithtexplus{figsbaladi3.ps}{baladi3.tex}{5.5}{12.9}{-0.3}{The
map $\Phi$ maps the left puzzle affinely onto the right puzzle,
respecting the shadings. Note
that horizontally, all domains are stretched (by $\lambda $) under the
map, while the vertical directions are squeezed (by $\lambda $). Also
note that the overall shape of the domain is that of the graph of $h$ of \fig{figsbaladi2.ps}}\cr
$$
\eqalign{
t(x)\,=\,\cases
{2\lambda ^2 ~,& if $x<{2\lambda ^2\over 1+\lambda }~,$\cr
1+\lambda ^2~,& if ${2\lambda ^2\over 1+\lambda }<x<{2\lambda
^2+3\lambda +1\over \lambda ^3+\lambda ^2}~,$\cr
0~,& if ${2\lambda
^2+3\lambda +1\over \lambda ^3+\lambda ^2}<x<{2\over \lambda ^2}~,$\cr
0~,&if ${2\over \lambda ^2}<x~.$}
}
$$
Since the first component of $\Phi$ is the 1-dimensional map $f$ we
discussed above, we see that correlation functions for observables
depending only on $x$ will show the resonances we found there. But the
map is uniformly contracting in the $y$ direction, and furthermore, we
have the explicit expression
$$
\Phi^n(x,y)\,=\, \left( f^n(x), {y\over \lambda ^n}+{\sum_{j=0}^{n-1}
t\bigl(f^j(x)\bigr)\over \lambda ^{n-j+1}}\right )~.
$$
If $F$ depends only on $x$ and is of the form $F(x,y)=u(x)$ we get
$$
\int dx\,dy\, F(\Phi^n(x,y))\, G(x,y)\,=\,\int dx \,u\big(f^{n}(x))v(x) dx~,
$$
where
$$
v(x)=\int dy\,G(x,y)~.
$$
Therefore, by the results of the preceding section, 
for generic $u$ and $G\in \CC^{1}$ we get a rate of decay of
correlations 
$\big|\lambda_{r,\pm}/\lambda\big|$. On the other hand, there is a
codimension two subspace of functions $G$
such that the decay rate drops down to $1/\lambda$.

\SECT{Baker}{Asymmetric Baker Map with Non-Trivial Essential
Decorrelation Radius}In
this section we give examples of maps whose
\EDR{}  for observables in
$\CC^1$
{\em larger} than $1/\lambda_{\min} $.
These maps $f_a$ are usually called asymmetric baker maps.
These are maps from
$[0,1]\times[0,1]$ which are defined as follows. Fix 
$a\in(0,1)$.
Then one defines
\figurewithtexplus{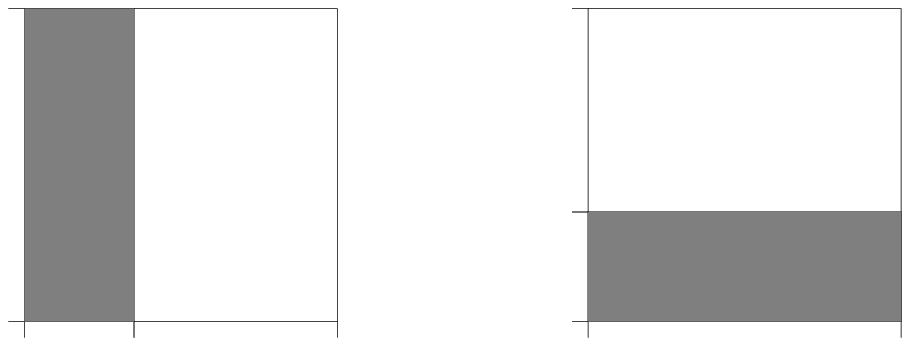}{baker1.tex}{6}{12}{-1.8}{An asymmetric
baker map. The gray rectangle on the left is mapped (preserving
orientation) affinely on the gray rectangle on the right. The white
rectangle is mapped on the white rectangle.}\cr
$$
f_a{x\choose y}\,=\,\cases
{\left (\matrix{{1\over a}& 0\cr
0 &a}\right )\left(\matrix{x\cr y\cr}\right )~,& if $0\le x\le a$~,\cr
\left (\matrix{{1\over 1-a}& 0\cr
0& 1-a}\right )\left(\matrix{x\cr y\cr}\right )+\left (\matrix{{-a\over 1-a}\cr a\cr}\right
)~,& if $a< x\le 1$~.\cr
}
$$
These maps have Jacobian equal to 1 everywhere, are invertible, and the
Lebesgue measure $\mu$ is the only absolutely continuous invariant
measure. The inverse is given by 
$$
f_a^{-1}{x\choose y}\,=\,\cases
{\left (\matrix{{a}& 0\cr
0 &{1\over a}}\right )\left(\matrix{x\cr y\cr}\right )~,& if $0\le y\le a$~,\cr
\left (\matrix{{ 1-a}& 0\cr
0&{1\over  1-a}}\right )\left(\matrix{x\cr y\cr}\right )+\left
(\matrix{a\cr
{-a\over 1-a}\cr }\right
)~,& if $a< y\le 1$~.\cr
}
$$
The Liapunov multipliers of $f_a$ are the exponentials
of 
$$
\mp\left (\mu(\{y<a\})\log a+\mu(\{y>a\})\log (1-a)\right )~,
$$
and so we find:
$$
\lambda_-\,=\,1/\lambda_+\,=\,a^a\cdot (1-a)^{1-a}\,\le\,1~.
$$
Note that with our notation, $\lambda_{\min}=\lambda_+$.

We next study the decay of the correlation functions.
Consider the two observables:
$$
F(x,y)\,=\,\partial _x u(x)~, \quad G(x,y)\,=\,x~,
\EQ{u} 
$$
with $u(0)=u(1)=0$, $u\ge0$, $u\not\equiv 0$. 
Note that since $F$ has zero average, it is not necessary to
impose that $G$ has zero average.
Then, with $z=(x,y)$ we find
$$
\eqalign{
S_k\,&=\,\int d^2z \, F\bigl(f^k(z)\bigr) G(z)\,=\, \int
d^2z\,F\bigl(z\bigr) G\bigl(f^{-k}(z)\bigr)\cr
\,&=\, \int
d^2z\,u'\bigl(x\bigr) G\bigl(f^{-k}(z)\bigr)~.\cr}
$$
When $y$ is fixed, we let $I_y$ be the horizontal segment $I_y=\{ (x,y)~:~
x\in[0,1]\}$. Note that $\left .  f^{-k}\right |_{I_y}$ is regular,
without discontinuities.
Therefore, we can integrate by parts and get
$$
S_{k}\,=\, -\int
d^2z\,u\bigl(x\bigr) \partial _x G\bigl(f^{-k}(z)\bigr)\cdot
\partial _x (f^{-k})_1(z)~.
\EQ{ip1}
$$
By construction $\partial _x G\equiv 1$ and
thus we find with $e_1\equiv{1\choose0}$:
$$
\eqalign{
S_k&=\,-\int d^2z \,u(x)\, \partial _x (f^{-k})_1(z)\,=\,
-\int d^2z \,u(x)\,  \left . D(f^{-k})\right|_{z}e_1\cr
\,&=\, 
-\int
d^2z\,u\bigl(x\bigr)\cr&~~~\cdot  \exp\kern -2.7pt\left ( \log a \cdot \sum_{j=0}^{k-1}
\chi_{y<a}\bigl(f^{-j}(z)\bigr)+\log (1-a) \cdot \sum_{j=0}^{k-1}
\chi_{y>a}\bigl(f^{-j}(z)\bigr)\right )~.\cr
}
$$
Note that  the exponential does not depend on the first component $x$
of $z\in\real^2$ and thus we can
integrate over $x$ and obtain
$$
S_k=
-\int
dy\,  \exp\kern -2.7pt\left ( \log a \cdot \sum_{j=0}^{k-1}
\chi_{y<a}\bigl(f^{-j}(z)\bigr)+\log (1-a) \cdot \sum_{j=0}^{k-1}
\chi_{y>a}\bigl(f^{-j}(z)\bigr)\right )~.
$$
By an explicit computation we see that the integral over $y$ equals
$$
\eqalign{
\sum_{j=0}^{k-1} &{k-1 \choose j} a ^j (1-a)^{k-1-j} \exp\bigl({j \log a +
(k-1-j)\log (1-a)}\bigr)\cr
\,&=\,\sum_{j=0}^{k-1} {k-1 \choose j} a ^j (1-a)^{k-1-j} a^j (1-a)^{k-1-j}\cr
\,&=\, \bigl(a^2+ (1-a)^2\bigr)^{k-1}~.
}
$$
\vskip -0.7cm
\figurewithtexplus{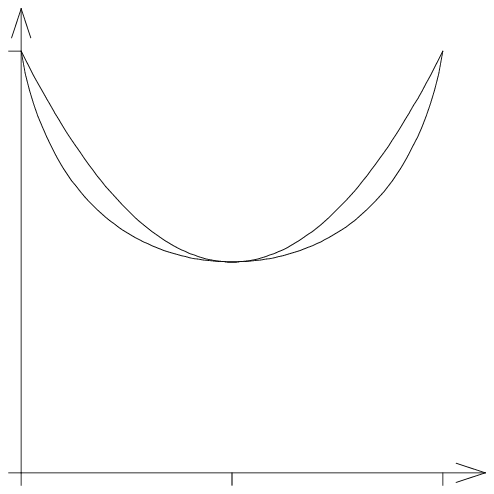}{baker2.tex}{6}{6}{0}{The
{\em upper} curve shows the
rates of decay $1/\lambda_{\min}=1/\lambdastar =a^2+(1-a)^2 $ as a function of $a$, and the
{\em lower} curve
shows $1/\lambda_+= a^a (1-a)^{1-a}$. One can see that the decay is
generally {\em 
slower} than $1/\lambda_+=1/\lambda_{\min}$.}\cr
At this point one needs to show that enough functions $F$ and $G$ have
been constructed to really characterize the \EDR. This follows by a
(simpler) application of the ideas of \sec{GC}. Note, however, that
the basic ingredient will still be the integration by parts formula \equ{ip1}. 
Leaving this problem aside, we get
$$
|S_k|\,\ge\, C \bigl(a^2+(1-a)^2\bigr)^k\,\equiv\, C \lambdastar  ^{-k}~,
$$
and
$$
1/\lambdastar  \,=\,a^2+(1-a)^2 \,\ge\, 1/\lambda_+\,=\, a^a
(1-a)^{1-a}~,
\EQ{CI000}
$$
{\em with equality only in the case of uniform expansion, $a=\HALF$
(and the identity maps $a=0$, $a=1$).}
Thus, the decay rate is {\bf not} given by the inverse of the
expanding Liapunov multiplier.

\SECT{vv}{An Essential Decorrelation Radius above $1/\lambda $ and a Resonance}In
this section, we 
somewhat generalize the construction of \sec{Baladi1} and give an
example of a map of the interval which has a resonance $\lambda_{\rm r}$ with
$|1/\lambda_{\rm r}|>1/\lambdastar $ and for which
also $1/\lambdastar >1/\lambda $. This map is obtained as a perturbation
of the Baladi map.

Consider four consecutive intervals $I_{1},\dots,I_{4}$ in increasing
order. We consider a map $f$ of $I=I_{1}\cup\cdots\cup I_{4}$ into
itself which is affine on each interval and satisfies the (topological) 
Markov property
$$
\eqalign{f(I_{1})\,&=\,I_{2}\cup I_{3}~,\cr
f(I_{2})\,&=\,I_{4}~,\cr
f(I_{3})\,&=\,I_{4}~,\cr
f(I_{4})\,&=\,I_{1}\cup I_{2}\cup I_{3}~.\cr
}
$$
We will denote by $l_{1},\dots,l_{4}$ the lengths of the intervals, and
by $f_{1},\dots,f_{4}$ the absolute value of the slope of $f$ in each
interval. In
order to ensure 
the above topological Markov property, some relations are required
between the lengths and the slopes, namely
$$
\eqalign{l_{1}f_{1}&=l_{2}+l_{3}~,\cr
l_{2}f_{2}&=l_{4}~,\cr
l_{3}f_{3}&=l_{4}~,\cr
l_{4}f_{4}&=l_{1}+l_{2}+l_{3}~.
\cr
}
\EQ{contraintes}
$$
In order to ensure the differentiability of the map at the
fixed point, we will assume that $f_{3}=f_{4}$. Note that the Baladi
map is the particular case when all slopes have equal modulus. 
If the slopes are given, the system \equ{contraintes} is composed 
of  four homogeneous equations in four
unknowns (the lengths). A necessary condition for the existence of a solution
is the vanishing of the determinant
of the associated matrix, namely
$$
f_{1}f_{2}f_{3}^{2}-f_{1}f_{2}-f_{1}f_{3}-f_{2}-f_{3}\,=\,0~.
\EQ{det}
$$
Note that if all slopes are equal to $\lambda$, then \equ{det} is
equivalent to 
the equation $\lambda ^4-2\lambda ^2-2\lambda =0$ for the Baladi map. 
We can also write the above relation as 
$$
f_{1}\,=\,{f_{2}+f_{3}\over f_{2}f_{3}^{2}-f_{2}-f_{3}}~.
$$
For any choice of $f_{2}$ and $f_{3}$ sufficiently close to $\lambda$, the
above expression defines a number $f_{1}$ again close $\lambda$ which
is therefore
larger than one. We can now choose $l_{4}>0$ and define
$$
l_{2}\,=\,{l_{4}\over f_{2}}~,\quad
l_{3}\,=\,{l_{4}\over f_{3}}~,\quad
l_{1}\,=\,{l_{2}+l_{3}\over f_{1}}~.
$$ 
The last equation of \equ{contraintes} is automatically satisfied and we
have an obviously positive solution for the set of lengths which can
be normalized to $\sum_i l_i=1$.

Having constructed our maps, we now investigate the
Perron-Frobenius (PF) operator on the 
set of functions which are piecewise constant on  the atoms
$I_{1},\dots,I_{4}$ of the topological Markov partition. These
functions are in bijection with four vectors, and it is easy to verify
that the  PF operator on these vectors is given by the matrix
$$
P\,=\,\pmatrix{0&0&0&{1\over f_{3}}\cr
{1\over f_{1}}&0&0&{1\over f_{3}}\cr
{1\over f_{1}}&0&0&{1\over f_{3}}\cr
0&{1\over f_{2}}&{1\over f_{3}}&0\cr
}~.
$$
The eigen-equation for this matrix is
$$
\xi^{4}-{\xi^{2}\over f_{3}}\left({1\over f_{2}}+{1\over f_{3}}\right)
-{\xi\over f_{1}f_{3}}\left({1\over f_{2}}+{1\over f_{3}}\right)=0~.
$$
It follows easily  from  relation \equ{det} that $\xi=1$ is a solution. 
Since $\xi=0$ is also a solution, and by continuity, for $f_{1},\dots,f_{3}$
near $\lambda$, we must have a resonance (close to $\lambda_{{\rm
r},\pm}$) given by the solutions of
$$
\xi^{2}+\xi+1-{1\over f_{3}}\left({1\over f_{2}}+{1\over f_{3}}\right)=0~.
$$
{\em Thus, we have constructed maps $f$ with both a resonance (when
the $f_i$ 
are close to $\lambda $ and chosen as indicated above) and with
non-constant slope.} From the discussion of \sec{CI} we will get
immediately
\CLAIM{Proposition}{bbb}{There is a piecewise affine map of the
interval which has a resonance, and for which the \EDR{}  
is {\bf larger} than $1/\lambda$, where $\lambda $ is the
Liapunov multiplier of the map (for the unique absolutely continuous
invariant measure).}

Furthermore, by continuity, when we are close enough to the Baladi
map, we find
$$
|1/\lambda_{\rm r}| \,>\, 1/\lambdastar   \,>\, 1/\lambda\,=\,1/\lambda_{\min} ~.
$$
\LIKEREMARK{Resonance and sub-optimal decay}{Using the above construction of
a map with a resonance and non-constant 
slope, one can also construct a skew product in a similar way as
in \sec{Baladiskew} and obtain a hyperbolic map {\em with a resonance}
and with decay which is {\bf slower} than $1/\lambda_{\min} $.}

\SECT{CI}{Maps of the Interval with Essential Decorrelation Radius above
$1/\lambda $}In this section we show that there are many maps of the
interval (or the circle) for which the \EDR{} is {\em
larger} than $1/\lambda $, where $\lambda $ is the Liapunov exponent
(for the absolutely continuous invariant measure).
Our results hold for maps with constant slope in each piece of the
Markov partition. 
They are based on the work of
Collet and Isola [\ref{ColletIsola1991}] who generalized the
inequality \equ{CI000} to 
more general 1-dimensional systems. For these there is
an explicit formula 
both for the Liapunov multiplier and the essential spectral radius.
For maps with constant slope in each piece the methods of
[\ref{ColletIsola1991}] can be generalized to show
that $\rhoess=\sigmaess$.
Therefore, the {\em equality between
$1/\lambda $ and the \EDR{}  only holds when the map has the same
(absolute value of the) slope {\bf everywhere}}.\myfoot{To some extent these formulas can
be generalized to hyperbolic SRB 
systems as we will show in \sec{GC}.}

The Liapunov multiplier for any invariant measure $\mu$ is given by
$$
\lambda_\mu \,=\,\exp \left (\int d\mu(y) \log |f'(y)|\right )~.
\EQ{ci999}
$$
One has also for almost every $x$ with
respect to the measure $\mu$ the more physical form
$$
\eqalign{
\lambda_\mu \,&=\,\lim_{n\to\infty }\left (\prod_{j=0}^{n-1}
|f'\bigl(f^j(x)\bigr)|\right )^{1/n}\,=\,
\lim_{n\to\infty }\exp\left (  {1\over n}\sum_{j=0}^{n-1} \log
|f'\bigl(f^j(x)\bigr)|\right )~.\cr
}\EQ{ci0}
$$
The identity between \equ{ci999} and \equ{ci0} is based on the
invariance and ergodicity of the measure $\mu$.
The results of [\ref{ColletIsola1991}] apply to 1-dimensional maps
with the following 
properties: 
There is a finite set of disjoint open intervals $I_1,\dots,I_\ell$ whose
closure forms a covering of $[0,1]$, and the closure of $f(I_k)$ is $[0,1]$
for every $k$.  We also
assume that $f$ is $\CC^{2}$ on each interval 
$I_{k}$ with a $\CC^{2}$ extension to the closure, and 
 $\tau <|f'|_{I_k}< \tau '$ with $\tau >1$.\myfoot{Note
that if all the slopes are positive, we are really talking about an
$\ell$-fold map of the circle to itself.} Ergodicity follows from the
previous assumption,  since  the Lebesgue measure of
$f^{-m}(I_j)\cup {I_i}$ is not zero for every $i,j$ and
any $m>0$. 
In this case, there is a unique absolutely continuous
invariant measure $\mu$, and using \equ{ci999} we will call $\lambda
=\lambda_\mu$ for this measure.

There is detailed information on the essential spectrum:
\CLAIM{Theorem}{CI}{{\rm[\ref{ColletIsola1991}]} The essential
spectral radius $\sigmaess$, where $\LL$ is the
Perron-Frobenius operator,
is given by
$$
\sigmaess(\CC^1,\LL)  \,=\, \exp\left(\lim_{n\to\infty }{1\over
n}\log\int d\mu(x)\, |(f^n)'(x)|^{-1}\right)~.
\EQ{ci2} 
$$
}

As we have seen in \sec{abc}, the relevant quantity is $\sigmapess$
and not $\sigmaess$. (See also [\ref{Keller1984}] for an early
reference.)
In [\ref{ColletIsola1991}], it was shown that
modulo a compact operator, each point of the open disk of radius
$\rhoess$ is an eigenvalue. Closer inspection of the argument used
there shows that for maps with {\em constant slope} in each piece of
the Markov partition the compact piece mentioned above has no effect,
since the boundary terms in [\ref{ColletIsola1991}, proof of Lemma 5]
do not contribute. Therefore, one finds
\CLAIM{Theorem}{CII}{The essential point
spectral radius  $\sigmapess$for maps with {\bf constant derivative in
each piece 
of the Markov partition} $\sigmapess$
is given by
$$
\sigmapess(\CC^1,\LL)  \,=\, \exp\left(\lim_{n\to\infty }{1\over
n}\log\int d\mu(x)\, |(f^n)'(x)|^{-1}\right)~.
\EQ{ci2a} 
$$
}

\CLAIM{Conjecture}{CIIII}{The identity \equ{ci2a} also holds for maps
with variable slope.}

\REMARK{For the maps of \clm{CII}, we therefore find the inequality:
$$
\rhoess(\CC^1,\CC^1,U)\,\ge\, \exp\left(\lim_{n\to\infty }{1\over
n}\log\int d\mu(x)\, |(f^n)'(x)|^{-1}\right)~.
\EQ{newer}$$
This means that for such maps the decay of correlations is indeed
related to the Liapunov multiplier, and as we shall see below in
\clm{conjugation}, equality only holds if all the slopes are the same
(in modulus).
}
\LIKEREMARK{Proof of \clm{CII}}We start with a setting which is
somewhat more general than the assumptions of \clm{CII}.
We consider a map $f$ of the unit
interval which is piecewise $\CC^{2}$ 
expanding and Markov, namely there is a finite partition $\cal A$ of the
interval by subintervals such that on each atom $f$ is monotone and
$\CC^{2}$ on the closure and such that the image of each atom is the union
of atoms (modulo closure). We also assume that there is an integer $k$
for which $|{f^{k}}'|>\zeta>1$, and $f$ is topologically mixing. Under
these assumptions it is well known that $f$ has a unique absolutely
continuous invariant probability measure $d\mu=h\;dx$ which is ergodic
with exponential decay of correlations (see [2] and references therein).
It is also easy to verify that $h$ is $\CC^{1}$ on each atom of $\cal A$
(with $\CC^{1}$ extension to the closure).

We will consider the decay of correlations in the space $X$ of functions
which are $\CC^{1}$ except maybe on the boundary of the atoms of ${\cal
A}$. For  $Y$, we use the space of $\CC^{1}$ functions whose
integral over each atom of $\cal A$ is equal to zero. This insures
that if $g\in Y$, we can find a function $v\in \CC^{2}$ such that $v'=g$
and $v$ vanishes on the boundary of the atoms of $\cal A$.

We will denote by ${\cal A}_{n}$ the partition
$\bigvee_{0}^{n}f^{-j}{\cal A}$. 
If $u$ and $v$ are $\CC^{1}$
functions, we have 
$$
\int u\cdot v'\circ f^{n} d\mu=
\int {u h\over {f^{n}}'}\cdot v'\circ f^{n} {f^{n}}' dx
=\sum_{I\in {\cal A}_{n-1}}
\int_{I} {u h\over {f^{n}}'}\cdot v'\circ f^{n}\cdot {f^{n}}' dx~,
$$
and integrating by parts we get
$$
\eqalign{
\int u\cdot v'\circ f^{n} d\mu\,&=\,
-\sum_{I\in {\cal A}_{n-1}}
\int_{I} \left({u h\over {f^{n}}'}\right)'v\circ f^{n} dx\cr
+\sum_{I\in {\cal A}_{n-1}}&{u(b_{I}^{-})h(b_{I}^{-})v(f^{n}(b_{I}^{-}))\over
{f^{n}}'(b_{I}^{-})}
-\sum_{I\in {\cal A}_{n-1}}{u(a_{I}^{+})h(a_{I}^{+})v(f^{n}(a_{I}^{+}))\over
{f^{n}}'(a_{I}^{+})}\;,}\EQ{unequation}
$$
where the boundary points $a_{I}$ and $b_{I}$ are defined for $I\in {\cal A}_{n-1}$ by
$$
\overline I=[a_{I},b_{I}]\;.
$$
Note that the two sequences $(a_{I})$ and $(b_{I})$ are identical except
for the first and last terms, and they are given by all the preimages of
order up to $n-1$ of the boundaries of the atoms of $\cal A$. In
particular, for each $I\in {\cal A}_{n}$, $f^{n}(a_{I})$ and $f^{n}(b_{I})$
belong to $\partial {\cal A}$.

We now use the assumption of \clm{CII}, namely that $f'$ is constant on the
atoms of $\cal A$. This implies that ${f^{n}}'$ is constant on the atoms
of ${\cal A}_{n-1}$. Therefore, the first term of \equ{unequation} is given by
$$
-\sum_{I\in {\cal A}_{n-1}}
\int_{I} \left({u h\over {f^{n}}'}\right)'v\circ f^{n} dx
\,=\,-\int{(u h)'\over {f^{n}}'}\;v\circ f^{n} dx
\,=\,-\int {\cal L}^{n}\left({(u h)'\over {f^{n}}'}\right)v dx~,
$$
where $\cal L$ is the Perron-Frobenius operator associated to $f$. Note
that when $f$ is not constant on each atom of $\cal A$, another term
appears involving the derivative of ${f^{n}}'$. This term corresponds to
a compact operator and did not intervene in the computation of the
essential spectral radius in [\ref{ColletIsola1991}]. 
It is not clear how such a term
would influence the present computation. 

To complete the proof of \clm{CII} one first applies \clm{conj1} to the
operator
$$
U(g)={\cal L}\left({g\over f'}\right)
$$
in the space $X'$ of functions which are piecewise $\CC^{0}$ except
possibly at the boundary of the atoms of $\cal A$, and $Y'$ the space of
$\CC^{2}$ functions vanishing on $\partial{\cal A}$.  One then  applies
Lemma 5 of [\ref{ColletIsola1991}] to conclude that each point in the
open disk of the essential spectrum is an eigenvalue. (This Lemma has
only been proven 
for full Markov maps 
but the proof easily extends to the general Markov case.) Note
that since $h\neq0$, multiplication by $h$ is a bounded invertible
operator in $X$. This
provides the desired lower bound if there is only the first term in
equation \equ{unequation}.

It remains to show that the last two terms are equal to zero, but this
follows at once from the requirement $v(\partial{\cal A})=0$.

\REMARK{The r.h.s.~of \equ{newer} is a special value of the function
$$
F(\beta )\,=\,\lim_{n\to\infty }{1\over n}\log\int d\mu(x)\,
|(f^n)'(x)|^{\beta }~,
$$
at $\beta =-1$. The function $F$ is convex, and its derivative at
$\beta = 0$ is the Liapunov exponent, by \equ{ci0} (for the measure
$\mu$ after exchanging limits and derivatives which can be justified in
that case):
$$
\eqalign{
\left . \partial _\beta \lim_{n\to\infty }{1\over n}\log\int d\mu(x)\,
|(f^n)'(x)|^{\beta }\right |_{\beta =0}\,&=\,
\lim_{n\to\infty }{1\over n}\log\int d\mu(x)\,
\log|(f^n)'(x)|\cr
\,&=\,
\log\int d\mu(x)\,
\log|f'(x)|\,=\,\log\lambda_\mu ~,\cr
}
$$
by the invariance of the measure $\mu$.
}

We next discuss the relation between $\lambdastar $ and $\lambda $.
The function $F$ is related to the pressure $P$ of the observable $-\log
f'$ (we refer to [\ref{Ruelle1978}] for the definition) by the relation
$$
F(\beta)=P((\beta-1)\log|f'|)~,
\EQ{pressure}
$$ 
since  $P(-\log|f'|)=0$ and $F$ is defined with respect to the SRB
measure $\mu$.
The quantity $1/\lambdastar $ is bounded from above by:
$$
1/\lambdastar \,\le\,\lim_{n\to\infty }  \sup_x
\left |{1\over \bigl(f^n\bigr)'(x)}\right |^{1/n}\,\le\,\sup_x |{1\over f'(x)}|~.
$$
And from below, it is bounded by $1/\lambda $, using Jensen's inequality
$$
\int d\nu(x)\exp\bigl(u(x)\bigr)\,\ge\, \exp\bigl(\int d\nu(x) u(x)\bigr)~,
$$
which holds for any probability measure $\nu$:\myfoot{The third line
uses again the invariance of the measure as in \equ{ci0}. }
$$
\eqalign{
-\log \lambdastar &=\lim_{n\to\infty }{1\over n}\log\int d\mu(x)\,
|(f^n)'(x)|^{-1}=\lim_{n\to\infty }{1\over n}\log\kern-0.2em\int d\mu(x)\,
\exp\left (-\log|(f^n)'(x)|\right )\cr
\,&\ge\,\lim_{n\to\infty }{1\over n}\log
\exp\left (-\int d\mu(x)\log|(f^n)'(x)|\right )\cr
\,&=\,\lim_{n\to\infty }{1\over n}
\left (-\int d\mu(x)\sum_{j=0}^{n-1}\log |f'\bigl(f^j(x)\bigr)|\right )\cr
\,&=\,-\int d\mu(x)\log |f'(x)|\,=\,-\log \lambda~. 
}\EQ{ci21}
$$
Thus, we see from \clm{CI} that $1/\lambdastar \ge 1/\lambda $.

On the other hand, if $|f'|$ is constant (and hence equal to $\lambda $), we
always have $\lambdastar =\lambda $, as one sees immediately from
\equ{ci2}. More interestingly, the converse holds as well, modulo conjugations:
\CLAIM{Theorem}{conjugation}{One has $\lambda =\lambdastar $ if and only
if there exists a $\Psi$ of bounded variation for which
$$
\Psi \circ f(x) \,=\,T_\lambda  \Psi (x)~,
\EQ{conjugate}
$$
where $T_{\lambda}$ is a map with piecewise constant slope $\pm
\lambda $.
Furthermore, such a $\Psi $ exists if and only if 
$$
\var(u)\equiv\lim_{n\to\infty}{1\over n}
\int\bigg(\sum_{j=0}^{n-1}u\circ f^{j}\bigg)^{2}\;d\mu~
\EQ{variance}
$$
vanishes for $u=\log|f'|-\log\lambda $.
}

Another way to say this is:
\CLAIM{Corollary}{x}{One has $1/\lambdastar >1/\lambda $ if and only if
$u=\log|f'| -\log\lambda $ {\bf fluctuates} in the sense that $\var(u)\ne0$.}

\LIKEREMARK{Proof of \clm{conjugation}}The anchoring point of the
proof will be the variance.
First of all, by
\refb{Keller1979} the limit in \equ{variance} always exists. 
Take now $u=\log |f'| -\log\lambda$, where $\lambda=\lambda_\mu $ is again
the Liapunov multiplier.
If  $\var(u)=0$, then
by a result of Rousseau-Egele [\ref{Rousseau-Egele1983}, Th\'eor\`eme
2, Lemme 6]
there exists a function $w$ of bounded variation such that $u=w\circ
f-w$ so that for our particular choice of $u$ one has:
$$
\log |f'| -\log\lambda \,=\,w\circ f -w~,
$$
and exponentiating
$$
|f'|e^{-w\circ f} \,=\,\lambda e^{-w}~.
\EQ{temporary}
$$
Note that $\exp(w)$ and $\exp(-w)$ are also
of bounded variation.
To keep the argument simpler, we will consider only the case when
$f'>0$ and work on the circle,
and leave the details of the case where the sign of $f'$ can
change to the reader. In this case we find from \equ{temporary} with
$\Psi(x)=\int_0^x ds\, 
e^{-w(s)}$ the identities
$$
\eqalign{
(\Psi\circ f)'\,=\,f' e^{-w\,{\circ}\, f}\,=\,\lambda e^{-w}\,=\,\lambda \Psi'~,
}
$$
and therefore $\Psi\circ f=T_\lambda \Psi $. So we conclude that if
$\var(u)=0$ the required $h$ exists, and furthermore, computing
\equ{ci21} in the coordinate system defined by $\Psi$, we see that
$\lambdastar  =\lambda $.

If $\var(u)>0$, then, since $F$ of \equ{pressure} is a convex function and 
$F''(0)=\var(\log |f'| -\log\lambda )$, we see
that $\lambdastar <\lambda $. 

Finally, if $\lambdastar <\lambda $ then clearly $f$ cannot be
conjugated to a function with constant slope, because in that case we would have
$\lambda =\lambdastar $ from \equ{ci21}.

This completes the proof of \clm{conjugation} (and also of \clm{x}).\QED

\SECT{GL}{Expanding Maps of Smooth Manifolds}The 
results of [\ref{ColletIsola1991}]
have been extended to the multi-dimensional {\em expanding} case in
the work of Gundlach and Latushkin
[\ref{GundlachLatushkin:2001}]. 
Simplifying their statement for our purpose, they show the following 
\CLAIM{Theorem}{GL}{The Perron-Frobenius operator for a $\CC^2$
expanding map $\phi$ of a smooth manifold $\MM$, when acting on  
the space of $\CC^1$
functions, has an essential spectral radius given by
$$
\sigmaess  \,=\, \exp \left (
\sup _{\nu\in {\rm Erg}}\bigg(h_\nu +\int _\MM \kern -0.5em d\nu(x)\,\log\bigl(
|\det {\rm D}\phi(x)|^{-1}\bigr) -\chi_ \nu\bigg)
\right )~,
\EQ{GL}
$$
where the sup is over all ergodic measures of the system, $h_\nu$ is
the entropy of the map w.r.t.~$\nu$ and $\chi_\nu$ is the smallest
Liapunov exponent of $D\phi$.\myfoot{This is obtained from Eq.(1.2) in
[\ref{GundlachLatushkin:2001}], where the authors allow a cocycle derived
from a bundle automorphism in
place of $D\phi$.} 
}

\REMARK{It should be noted that the hypotheses of \clm{GL}, in
particular the differentiability {\em everywhere} imply the existence
of a finite Markov partition for the map. It seems that no general
result is known in the absence of this condition.}

Before we use \equ{GL} in more general
contexts, we first show that one recovers indeed the formulas of
\clm{CI} when one considers the case of an
expanding map $f$ of the circle. In that case, one takes $\phi=f$.
For an invariant ergodic measure $\nu$ the integral in \equ{GL} equals
$$
\lambda_\nu\,=\,-\int d\nu\,\log|f'|~.
$$
The unique Liapunov exponent of $Df$ for the invariant ergodic measure
$\nu$ is 
$$
\chi_{\nu}\,=\,\int\log|f'|d\nu~.
$$
From \equ{GL}  we conclude the that
$$
\sigmaess \big(\CC^{1}\big)\,=\,
\exp\left ({\sup_{\nu\in{\rm Erg}}
\{h_{\nu}-2\int\log|f'|d\nu\}}\right )~.
$$
On the other hand, by the variational principle (see Ruelle
\refb{Ruelle1978}) we have
$$
\sup_{\nu\in{\rm Erg}} \{h_{\nu}-2\int\log|f'|d\nu\}
\,=\,P(-2\log|f'|)~.
$$
By \equ{pressure}, we have $P(-2\log|f'|)=F(-1)$,
which is \equ{ci2}, as asserted.\QED

We now consider the more general examples
covered by \clm{GL} and show that they indeed imply the same kind of
lower bound.
Note that by Ruelle's identity [\ref{Ruelle1978}] one knows that the
spectral radius $\sigmasp$ of the Perron-Frobenius operator on $\CC^{0}$ (or
$\CC^1$ if the transformation is regular enough, 
since it equals the maximum  positive eigenvalue) is
$$
\sigmasp \,=\,\exp \left (\sup _{\nu\in {\rm Erg}}\bigg(h_\nu -\int _\MM \kern -0.5em d\nu(x)\,\log
|\det {\rm D}\phi(x)|\bigg)\right) ~.
\EQ{ruelle}
$$
When $\mu$ is the sole invariant measure which is absolutely
continuous w.r.t.~Lebesgue measure 
we see by the variational principle \refb{Ruelle1978}
that the argument of the exponential is the pressure and hence
$$
\sigmasp \,=\,\exp \left (h_\mu -\int _\MM \kern -0.5em d\mu(x)\,\log
|\det {\rm D}\phi(x)|\right)\,=\exp{P(-\log|{\rm detD}\phi|)}=
\,\exp(0)\,=\,1 ~.
\EQ{ruelle2}
$$
To get a lower bound on the essential spectral radius, we can plug in a
particular measure in expression \equ{GL}. Using the  SRB
measure $\mu$, we get 
$$
\sigmaess \,\ge\, e^{-\chi_{\mu}}~.
$$
Thus, we get in this case the following corollary from \clm{GL}:
\CLAIM{Corollary}{CGL}{The essential spectral radius of the
Perron-Frobenius operator for a\ \ $\CC^2$ 
expanding map $\phi$ of a smooth manifold $\MM$ acting on the space of
$\CC^1$ 
functions satisfies
$$
\sigmaess \,\ge\, e^{-\chi_{\mu}}~,
$$ where $\chi_\nu$ is the smallest
Liapunov exponent of $D\phi$.}

\CLAIM{Question}{difficult}{The relation with $\rhoess$ remains open.}

\SECT{GC}{A Conjecture and Some Steps Toward its Proof}The setting is now
that of a smooth compact Riemannian
manifold $\MM $ and a uniformly hyperbolic 
diffeomorphism $f$ which is topologically mixing on the global
attracting set
$\Omega $.
We denote by $\mu$ the
unique SRB measure (see \refb{KatokHasselblatt1995}). We assume that the smallest
Liapunov multiplier larger than 1  is associated with a space of
dimension one.\myfoot{This means that
the smallest positive Liapunov exponent is associated with a space of
dimension one.}
Let $g_{1}$ and $g_{2}$ be two
observables whose regularity will be fixed below. Let ${\cal A}$ be a
Markov partition of $\Omega $, which is fine enough so that each atom can be foliated
by {\em local} stable and unstable manifolds (see
\refb{KatokHasselblatt1995}). From now on when we
speak of a local stable or unstable leaf, we always mean its restriction
to an atom of the Markov partition. When speaking of a function on an
atom $A_0$ of $\AA$, we mean a function on the corresponding rectangle
(the hull) on the ambient space $\MM$.

We can write the correlation function
$$
S_{k}\,=\,\int_{\MM }d\mu\,g_{1}\cdot g_{2}\circ \map^{k}\,=\,
\int_{\MM }d\mu\,g_{1}\circ \map^{-k}\cdot g_{2}\,=\,
\sum_{A\in {\cal A}}\int_{ A}d\mu\,g_{1}\circ \map^{-k}\cdot
g_{2}~.
\EQ{correlation}
$$

We begin by rewriting
\equ{correlation}  using the disintegration of the SRB measure with
respect to the unstable 
foliations (see \refb{KatokHasselblatt1995}). In other words, there is a measure $N$ on the set
$\WW ^{u}$ of local unstable leaves, and  for any $W\in\WW ^{u}$
there is a H\"older continuous positive function $\Theta_{W}$ on $W$ such
that
$$
S_{k}=\sum_{A\in {\cal A}}\int_{A\cap\WW ^{u}}dN(W)\int_{W}dM_W\,
g_{1}\circ \map^{-k}\cdot g_{2}\cdot\Theta_{W}~,
\EQ{hyp1}
$$
where $dM_{W}$ is the Riemann  measure on $W$, and where $A\cap
\WW ^{u}$ is the subset of elements of $\WW ^{u}$ contained in $A$.
We finally define the density $\h $ on the leaves by
$$
\h (x)=\Theta_{W(x)}(x)~.
\EQ{rhodef}
$$
Note that the function $\h $ may not be defined on the whole phase
space if we have a non trivial attractor (for example a strange
attractor). However one can interpolate this function to a globally
defined (strictly positive) H\"older continuous function, see \eg\
[\ref{Mattila1995}]. 
\REMARK All our problems are related to this density\myfoot{Note that
this density can be rough even if the invariant measure is the
Lebesgue measure}, because, as one
can see from \equ{hyp1}, the effective observable is not $g_2$ but
$g_2\cdot \h $, and therefore smoothness requirements on $g_2$ alone do not
suffice to make $g_2\cdot \h $ smooth enough.

Let $\delta $ be a positive constant whose value may
vary with the context and system. 
By
$\CC^{1+\delta }$ we mean the class of $\CC^1$ functions whose
derivative is $\delta $-H\"older continuous.

\CLAIM{Assumption}{hyp}{The foliation
$\WW^u$ by the local manifolds $W^{u}_{\rm
loc}$ is a $\CC^{1+\delta }$ foliation of $\CC^{1+\delta }$
manifolds and  
the field of one  dimensional directions   corresponding to the smallest
expanding direction is H\"older continuous.
Furthermore, $\h $ extends to a  H\"older
continuous function
on $\MM$ and $\CC^{1+\delta }$ in the unstable directions.}

We will need further assumptions on this foliation, see \fig{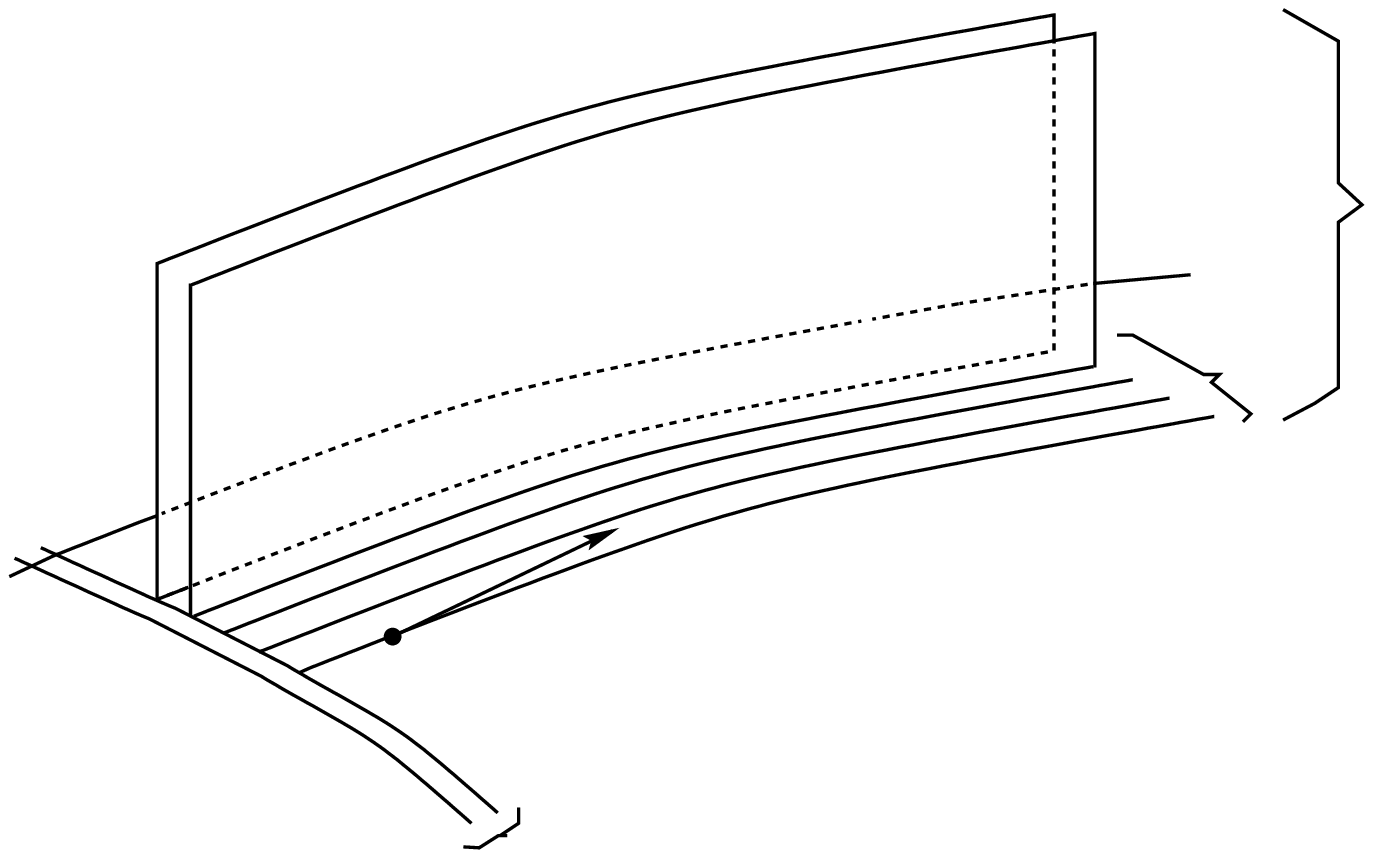}:
Denote by $\vec t_x$ the normalized tangent vector to $W^{u_{\min}}(x)$ at
$x$, where  $W^{u_{\min}}(x)$ is the (one-dimensional) manifold at $x$
corresponding to the slowest expanding direction. Note that by our
assumption it is H\"older in $x$.  
\figurewithtexplus{figswww.eps}{www.tex}{9.2727}{15}{1.0}{The
stable foliation ${\cal W}^s$ (of dimension 1) and the unstable 
foliation ${{\cal 
W}^u}$ (of dimension 2). Inside ${{\cal 
W}^u}$ lie the one-dimensional leaves $\{w\}$ of the most slowly expanding
direction, forming the family ${\cal W}^{u_{\rm min}}$. A leaf of
this family is called $w$, with a ``coordinate origin'' $x_w$, (where
$s=0$), and
the
tangent vector $\vec t_{x(w,0)}$ (here shown at $s=0$).
We also labeled two
leaves
$W^s$ and $W^u$ corresponding to the respective foliations $\WW^s$
resp.~$\WW^u$.}\cr

Since the field of vectors $\{\vec t_x\}$ is covariant,
we find (see \refb{KatokHasselblatt1995}) that there is a H\"older
continuous 
function $\varphi$ (defined on $\MM$) such that 
$$
Df_{x}\cdot \vec t_x=e^{\varphi(x)}\vec t_{f(x)}~.
\EQ{varphi}
$$
The function $\varphi$ (whose average is positive) is the ``local
expansion rate'' in the least unstable 
direction. 
Similarly, there is a H\"older
continuous differential 1-form $\alpha$ such that for any $x$
$$
\alpha_{f(x)}Df_{x}=e^{\varphi(x)}\alpha_{x}~,
\EQ{a1}
$$ 
and 
$$
\alpha\big(\vec t\;\big)_{x}\,\equiv\, \alpha_{x}\big(\vec
t_{x}\big)=1~.
\EQ{a2}
$$

To make the argument more transparent, we will pursue it for the case
of only 2 positive Liapunov exponents and leave the general case (with
heavier notation) to the reader. We have already fixed a tangent field
$\vec t$ and a 1-form $\alpha $ which measure what happens in the
``slow'' unstable direction. Similarly, we now introduce a tangent
field $\vec s$ and a 1-form $\beta $ which describe the other
unstable direction. These are unique and H\"older continuous.
The analogs of \equ{varphi}--\equ{a2} are then
$$
\eqalign{
Df_{x}\cdot \vec s_x\,&=\,e^{\eta(x)}\vec s_{f(x)}~,\cr
\beta_{f(x)}Df_{x}\,&=\,e^{\eta(x)}\beta_{x}~,\cr
\beta\big(\vec s\;\big)_{x}\,&\equiv\, \beta_{x}\big(\vec
s_{x}\big)=1~.\cr
}
$$

\CLAIM{Assumption}{ass3}{There are constants $\epsilon _{*}>0$ and
$C_*$ for which 
$$
\sum_{j=0}^{k-1}\eta\big(\map^{-j}({x})\big)\,\ge\,
\sum_{j=0}^{k-1}\varphi\big(\map^{-j}({x})\big)+k\epsilon _{*}+\log C_*~,
\EQ{xyzz}
$$
uniformly for sufficiently large $k$ and for  {\bf all} $x$ in $\MM$.}

This assumption implies $\alpha (\vec s\;)=\beta (\vec t\;)=0$,
because the Liapunov multipliers are different.
In
other words, the expansion rates $\eta$ 
and $\varphi$ are allowed to fluctuate, but there must remain a
``gap'' $\epsilon _{*}$ between them everywhere, and at large times. It
would be interesting to understand to which extent \equ{xyzz} could be
replaced by a condition on the Liapunov exponents alone. 
A stronger statement than \equ{xyzz} is to assume $\eta(x)>\varphi(x)$ for
all $x$. This
is in fact the ``bunching condition,'' since from the continuity of
$\varphi$ and $\eta$ and the compactness of
the manifold it follows that $\eta/\varphi>1+\epsilon >1$, uniformly
in $x$.

\REMARK This same condition ensures the H\"older continuity of the vector
field tangent to
$W^{u_{\min}}$
{\it i.e.}, it establishes one of the requirements
of \clm{hyp}. 

\REMARK The \clm{ass3} should be compared to the usual hyperbolicity
conditions [\ref{HPS1977}]. In that case, one requires for the stable
directions a 
bound of the form
$
\left . D(f^n)\right |_{E^s}\le \lambda_-^n
$
(in an adapted metric), and $\left . D(f^{-n})\right |_{E^u}\le \lambda
_+^{-n}$
for the unstable directions, and then $\epsilon _{*}=\log\lambda
_+-\log\lambda_-$. So for hyperbolicity, the strong form of \equ{xyzz}
is being required.

The following result is formulated as a conjecture, since the
arguments toward its proof are only sketched.

\CLAIM{Conjecture}{C1}{Consider a dynamical system which is uniformly
hyperbolic and 
has an SRB measure $\mu$
whose Liapunov multipliers are all different from 1.
Assume it satisfies \clm{hyp} and \clm{ass3}.
For observables which are piecewise\myfoot{In fact, the class of
observables we really consider is quite complicated, as it will turn
out to be a complicated subset of $\CC^1$. See below for a precise
description.}
$\CC^1$ in the unstable direction,
and Lipschitz continuous in the stable directions, the
\EDR{} is at least $1/\lambda_{\min}$,
where $\lambda_{\min}$ is the smallest Liapunov multiplier greater
than 1 (as in Eq.\equ{no2}).
Furthermore,
the essential spectral radius is strictly larger than  $1/\lambda
_{\min}$ whenever the system is not smoothly\myfoot{For more subtle
aspects of the conjugation, see \clm{conjugation}.} conjugated to a
system whose differential is a constant function in the
direction corresponding to $\lambda_{\min}$.} 

\CLAIM{Question}{remark}{We expect the conclusions to hold for
observable which are piecewise $\CC^1$ in all directions.
}

\LIKEREMARK{Sketch of proof of \clm{C1}}We first define the spaces $X$ and $Y$ for which can prove the
assertion of \clm{C1}. 
The space $X$ is formed by functions obtained as follows:
Select an atom $A_0$ of the partition $\AA$.
Choose a fixed vector field $\V_{A_0}$ which is defined in a neighborhood of $A_0$
and which is
tangent at every  $x\in A_0$ to $W^{u}_{\rm loc}(x)$, and which does
not vanish on the hull of $A_0$.
This is possible because
these manifolds are $\CC^{1+\delta}$ 
in $x$. We also assume that this vector field
has zero divergence in the unstable directions.

We further choose a function $v$ which is $\CC^{2}$ in the hull (in
$\MM$) of
$A_0$,
vanishing on the
stable boundaries of $A_0$.

The observables $g_2$ in $X$ are defined by the equation
$$
g_{2}\,=\,{1\over\h }dv(\V_{A_0} )~,
$$
where $h$ is defined in \equ{rhodef}, and is extended to a positive
function on the hull of $A_0$.
By our above assumptions, $g_{2}$ is $\CC^{1}$ in the
unstable directions and $\delta $-H\"older continuous in the stable
directions. As we vary $A_0$ and $v$ over all possible choices, we
obtain a set $X_0$ of functions. Since $v\in\CC^2$, the map $v\mapsto dv(\V_{A_0})$ has
closed kernel, and this induces a topology on the image. If we divide
by $h$, things do not change, and we have a topology on the functions
$g_2$. Varying $A_0$ this construction makes $X_0$ to a Banach space
$X$. (This space has a topology which is somewhat finer than the
$\CC^{1-\delta} $-H\"older topology considered above.)

We next construct the space $Y$. 
Fix an atom $A_0$. Let $j$ be a function on the hull of $A_0$ which is
$\CC^0$ along the unstable directions and $\delta $-H\"older in the
stable ones. Define $g_1$ by the equation
$$
dg_1(\vec t\,)_x\,=\, j(x)~.
\EQ{g1def}
$$
Arguing as in the construction of $X$, we obtain $Y$ by varying $A_0$
and $j$, and inducing the topology.

Finally, the operator $U$ is the Koopman operator of the map $f$, that is
$$
\bigl(U g\bigr)(x) \,=\,g\bigl(f(x)\bigr)~.
\EQ{Koop}
$$

Now that the spaces are in place, we can work on \equ{hyp1}. Take
$g_1$ and $g_2$ in a piece $A_0$ of the partition. 
We integrate by parts in \equ{hyp1} in each $W$ separately.
Since $\V $ is divergence-free and $v$ vanishes at the stable boundary
of $A_0$, we obtain 
$$\eqalign{
\int_{W}dM_Wg_{1}\circ f^{-k}\cdot g_{2}\cdot\Theta_{W}\,&=\,
\int_{W}dM_Wg_{1}\circ f^{-k}\,\cdot dv(\V )\cr
\,&=\,-\int_{W}dM_W\,d\big (g_{1}\circ f^{-k}\big)(\V )\cdot v
\cr\,&=\,
-\int_{W}dM_W\,\bigl({dg_{1}}\bigr)\circ{f^{-k}}\cdot
\big(D_{{x}}f^{-k}(\V )\big)\cdot 
v~.
}\EQ{intpart}
$$
Decomposing in the unstable directions we get
$$
\eqalign{
{dg_{1}}_{f^{-k}({x})}
\big(D_{{x}}f^{-k}(\V)\big)\,&=\,
dg_{1}(\vec t\;)_{f^{-k}({x})}\cdot
\alpha(\V )_{{x}}\cdot
e^{-\sum_{j=0}^{k-1}\varphi\big(\map^{-j}({x})\big)}\cr
\,&+\,
dg_{1}(\vec s\;)_{f^{-k}({x})}\cdot
\beta(\V )_{{x}}\cdot
e^{-\sum_{j=0}^{k-1}\eta\big(\map^{-j}({x})\big)}~.\cr
}
\EQ{gcond}
$$

\noindent Clearly, \clm{ass3} implies a uniform bound for \equ{gcond}:
$$
\eqalign{
{dg_{1}}_{f^{-k}({x})}
\big(D_{{x}}f^{-k}(\V )\big)\,&=\,
dg_{1}(\vec t\;)_{f^{-k}({x})}\cdot
\alpha(\V )_{{x}}\cdot
e^{-\sum_{j=0}^{k-1}\varphi\big(\map^{-j}({x})\big)}\cr
&~~~+\OO\bigl(e^{-\epsilon _*k
-\sum_{j=0}^{k-1}\varphi\big(\map^{-j}({x})\big)}\bigr)~,
}
\EQ{gcond2}
$$
so that the faster rate $\eta$ has been eliminated from the discussion.
Inserting in \equ{intpart}, we find 
$$
S_{k}\,=\,-\int d\mu\,
dg_{1}(\vec t\;\bigr)\circ \map^{-k}\,\,
e^{-\widetilde\Sigma_{k}}\tilde g_{2}+\OO(e^{-\epsilon _*}Z_{k})~,
\EQ{b1} 
$$
(as $k\to\infty $), where
$$
\displaylines{
\widetilde\Sigma_{k}(x)\,=\,\sum_{j=0}^{k-1}
\varphi^u\big(\map^{-j}(x)\big)~,\qquad
\tilde g_{2}(x)\,=\,{v(x)\over \h (x)}~,\qquad Z_{k}=\int d\mu\,e^{-\widetilde\Sigma_{k}}~,
}
$$
by the invariance of the measure, we find
$$
S_{k}\,=\,-\int d\mu \,\,\tilde g_{1}\,\,
e^{-\Sigma_{k}}\,\,\tilde g_{2}\circ \map^{k}+\OO(e^{-\epsilon _*}Z_{k})~,
\EQ{abc}
$$
where
$$
\Sigma_{k}(x)\,=\,\sum_{j=0}^{k-1}\varphi^u\big(\map^{j}(x)\big)~,\qquad
\tilde g_1(x) \,=\, dg_1(\vec t\,)_x~.
$$

We now use the thermodynamic formalism \refb{Bowen1975}.
Define the  H\"older continuous function $\varphi^{u}$ by
$\varphi^{u}=-\log \det J^{u}$ where
$J^{u}$ is the Jacobian matrix in the unstable bundle.\myfoot{In principle,
for the case of 2 positive Liapunov exponents,
$\varphi^u$ can be computed from $\varphi$, $\eta$, $\vec t$, $\vec
s$, $\alpha $, and $\beta $.}
Recall that under our assumptions there is a
homeomorphism $\pi$ conjugating the dynamical system $f$ on the
attractor to a subshift ${\cal S}$ of finite type. The SRB
measure $\mu$ is then
transformed to a Gibbs state $\gamma$ with the H\"older continuous potential
$\varphi^{u}\circ\pi$. 
We get for the first term on the right
hand side of \equ{abc}: 
$$
D_k\,=\,-\int d\gamma\;\tilde g_{1}\circ\pi\; e^{-\Sigma_{k}\circ\pi}\,
\cdot\, \tilde g_{2}\circ\pi\circ\S^{k}\;. 
\EQ{x1}
$$
Define an operator $\T$ by
$$
\T \psi\,=\,e^{-\varphi^u\circ \pi} \tilde \psi\circ \S~,
$$
where $\psi $ is a function on the shift space. Then \equ{x1} becomes
$$
D_k\,=\,-\int d\gamma\;\bigl(\T^*\bigr)^k (\tilde g_{1}\circ\pi)\; 
\cdot\, \tilde g_{2}\circ\pi\;. 
\EQ{x2}
$$
We now apply \clm{conj1} with $X'=Y'=\CC^0$ and $U'=\T^*$ and we find
$\rhoess(X,Y,U)=\rhoess(X',Y',U')\ge
\sigmapess(X',U')$. It remains to give a lower bound on $\sigmapess(X',U')$
in terms of the pressure. Using a well-known device [\ref{Baladi2000},
Lemma 1.3], we can conjugate $\T$ to an operator $\T_+$ defined by
$$
\T_+ \psi\,=\,e^{-\varphi_+^u\circ \pi} \tilde \psi\circ \S~,
$$
where $\varphi_+^u\circ \pi$ depends only on the future, \ie,
$\varphi_+^u$ is constant on the stable (local) leaves.
Note now that when $\T_+^*$ acts on a function $\psi_+$ which depends
only on the {\em future}, it is given by
$$
\T_+^* \psi_+\,=\,{1\over \phi_+} \LL\bigl(\phi_+e^{-\varphi_+^u\circ
\pi}\psi_+\bigr)~,
\EQ{almost}
$$
where $\LL$ is the Perron-Frobenius operator and $\phi_+$ satisfies
$$
\LL \phi_+\,=\,\phi_+~.
$$
The eigenvalue above is 1 because we are dealing with
an SRB measure, and the eigenvector is unique. We now see that 
$\sigmapess(X',U')$ is bounded below by the essential point spectral radius
of $\LL \exp({-\varphi_+^u\circ
\pi})$. One now introduces the pressure
$$
P(\varphi^u-h)\,\equiv\, \lim_{k\to\infty }{1\over k}\log\int_\MM d\mu(x)
\,e^{-\sum_{j=0}^{k-1}h\big(f^{-j}(x)\big)}~,
$$
where $P(\varphi^u)=0$, because we are dealing with an SRB measure. 
Then, it is known [\ref{ColletIsola1991}],
[\ref{Baladi2000}, Theorem 1.5.7] that every point in the open disk of
radius $\exp\bigl(P(2\varphi^u)\bigr)$ is an eigenvalue of $\LL \exp({-\varphi_+^u\circ
\pi})$. Since $\exp\bigl(P(2\varphi^u)\bigr)>1/\lambda_{\min}$, the desired
inequality follows. This completes the sketch of the proof of \clm{C1}.\QED

\SUBSECTION{Sufficient Conditions}We next address the question of sufficient conditions
for  \clm{hyp} and \clm{ass3} to hold.
A typical such condition is the bunching condition from
[\ref{KatokHasselblatt1995}, p. 602], 
or, the concept of domination developed in [\ref{HPS1977}].
Consider a point $x$ and write
$D\map_x$ in matrix form
$$
Df_x \,=\, \left (\matrix{ A_x& 0 \cr 0 &D_x\cr}\right )~,
$$
with the blocks corresponding to unstable and stable subspaces,
respectively. We define
$$
\lambda_x \,=\, \|D_x\|~,
\quad
\mu_x\,=\,\bigl(\|A_x^{-1}\|\bigr)^{-1}~.
$$
Let $\nu_x$ be the inverse of the Lipschitz constant for
$f^{-1}$:
$$
\nu_x\,=\,{1\over L(f^{-1})_x}~,\quad
L(g)_x\,\equiv\,\sup_{|x-y|<\epsilon  }{|g(x)-g(y)|\over |x-y|}~,
$$
with $\epsilon  >0$ some small constant.

One defines the bunching constant by
$$
B^u(f)\,=\,\inf_x{\log\mu_{x}-\log\lambda_{x}\over\log\nu_{x}}~.
$$

\CLAIM{Theorem}{smooth}{Let $\map$ be a $\CC^{3}$ map of the manifold
$\MM$ which 
gives rise to an Axiom A system whose unstable manifolds
$W^{u}$ are $\CC^{2}$ and the stable ones, $W^{s}$, are $\CC^{1}$. Assume that the Liapunov multipliers
of $\map$ satisfy the following conditions:
\item{1)}The smallest Liapunov multiplier above 1 is $\lambda_{\min}$ and the
corresponding dimension is 1.
\item{2)}There are no Liapunov multipliers equal to 1.
\item{3)}The (multidimensional version of) inequality \equ{xyzz} holds.
\item{4)}The bunching constant satisfies
$$
B^u(f)\,>\,1~,
\EQ{bunch1}
$$
and for the inverse map
$$
B^u(f^{-1})\,>\,1~.
\EQ{bunch2}
$$ 
Then \clm{hyp} and \clm{ass3} hold.
}

\PROOF{The proofs of all assertions except for the smoothness of $\h $
can be found in [\ref{KatokHasselblatt1995}, Chapter 19, p607].

So it remains to prove the differentiability of $\h $.
We recall that using a base point $x_{W}$ on the leaf $W$,
we have for the
density of the SRB measure on the unstable manifold $W^u(x)$ of any
$x\in W$:
$$
\h (x)=\prod_{j=0}^{\infty}e^{\varphi^{u}\bigl(\map^{-j}(x)\bigr)
-\varphi^{u}\bigl(\map^{-j}(x_{W})\bigr) }
$$
where $x_{W}$ is a reference point chosen once and for all on
$W^u(x)$.  When varying $x$ along an unstable leaf, the reference point
does not change. Each term in the above product is differentiable in
the unstable direction, and the regularity properties of $\h $ follow
easily by checking the convergence of the series. The H\"older
continuity of $\h $ follows by standard arguments from the  H\"older
continuity of the stable foliation. \QED 

\REMARK{In the case of skew products of Baladi maps, it is easy to
verify that the local stable manifolds are vertical segments. Because of
the local flatness of the invariant measure of the one dimensional
system, it follows from the explicit expression of the map that the
differential is diagonal. This implies that the field of unstable
directions is horizontal (as well as the local unstable manifolds). This
implies that $\Theta_{w}$ is constant on the local unstable manifold. By
changing if necessary the transverse measure, we can assume that
$\Theta_{w}=1$. Therefore, if $v$ is a $\CC^{2}$ function with compact
support contained in an atom of the Markov partition, the observable
$g_{2}$ defined as above is $\CC^{1}$ and we can apply the above
technique. Note that in this case \clm{hyp} is
violated since the map is area preserving.}

\SUBSECTION{An Example}We construct an example where all the
above assumptions are 
satisfied. This example is a generalization of the solenoid and can
also be viewed as a skew product. First of all, let $p>6$ be an odd
integer.  Let $\ell$ be that solution of
the equation
$$
\ell^{2}\bigl(1+\cos(2\pi/p)\bigr)-4\ell+2\,=\,0~,
$$
which is less than one. Let $r=1-\ell$. Note that for $p$ large, we have
$r\approx \pi/p$. Let $q$ be another odd integer with $p>q>4$.
It is easy to verify that the 
spheres of radius $r$ centered at points with polar
coordinates $(\ell,2k\pi/p,2m\pi/q)$ with $0\le k<p$ and $0\le m<q$
are mutually
disjoint.  We now define a map $f$ of ${\bf M}={\bf T}^{2}\times {\bf B}^{3}$
into itself (${\bf T}^{2}$ the two dimensional 
torus and ${\bf B}^{3}$ the three
dimensional unit ball) by
$$
f(\vartheta,\varphi,(x,y,z))\,=\,(p\vartheta,q\varphi, rx+\ell\cos\vartheta\cos\varphi,
ry+\ell\sin\vartheta\cos\varphi,rz+\ell\sin\varphi) ~,
$$
where the angles are modulo $2\pi$ and we use Cartesian coordinates on
${\bf B}^{3}$. It is left to the reader to verify that because of our
choice of $\ell$ and $r$ the map is injective. It is obviously a skew
product above the map of the torus $(\vartheta,\varphi)\mapsto
(p\vartheta, q\varphi)$ which is ergodic and mixing for the Lebesgue
measure. 
\REMARK More balls can be packed and also balls with larger radius using
a Peano surface for the position of the centers instead of the sphere of
radius $\ell$ as above.

The differential of $f$ is given by
$$
Df=\pmatrix{
p&0&0&0&0\cr
0&q&0&0&0\cr
{\cal X}&{\cal X}&r&0&0\cr
{\cal X}&{\cal X}&0&r&0\cr
{\cal X}&{\cal X}&0&0&r\cr
}~,
$$
and 
$$
Df^{-1}=\pmatrix{
1/p&0&0&0&0\cr
0&1/q&0&0&0\cr
-{\cal X}/(rp)&-{\cal X}/(rq)&1/r&0&0\cr
-{\cal X}/(rp)&-{\cal X}/(rq)&0&1/r&0\cr
-{\cal X}/(rp)&-{\cal X}/(rq)&0&0&1/r\cr
}~,
$$
where ${\cal X}$ denotes various quantities of order one. 

We now verify the bunching conditions. First of all, the stable
bundle is obviously obtained by setting the first two components of a
tangent vector equal to zero. Therefore, $\lambda=r$, and also the stable
manifold of a point $(\vartheta,\varphi,x,y,z)$
is the set of points with the same angles $\vartheta$ and $\varphi$. 

The unstable bundle is not so trivial. As in [\ref{KatokHasselblatt1995}], the
unstable bundle is obtained as a graph above the space of vectors whose
last two coordinates are equal to zero. In other words, for every point
$P\in{\bf M}$, there is a linear operator $L_{P}$ from ${\bf R}^{2}$ to
${\bf R^{3}}$ such that the unstable subspace at $P$ is the set
$$
E^{u}(P)=\left\{(z,L_{P}z)\,\big|\,z\in{\bf R}^{2}\right\}~,
$$
with the canonical identifications.  From the equation satisfied by
$L_{P}$ (see  [\ref{KatokHasselblatt1995}]) it follows easily that
$$
\sup_{P\in{\bf M}}\|L_{P}\|\,\le\,\Oun q^{-1}~.
$$
It then follows that $\mu^{-1}=q^{-1}(1+\Oun q^{-1})$. Finally
$\nu^{-1}$ is at most the sup norm of $Df^{-1}$ and we get $\nu^{-1}\le
r^{-1}+\Oun r^{-1}q^{-1}$. Recalling that $r\approx \pi q^{-1}$ for
large $q$, we get
$$
\lambda\mu^{-1}\nu^{-2}\,\le\, r q^{-1} r^{-2}(1+\Oun q^{-1})\,\le\,
\pi^{-1}(1+\Oun q^{-1})\,<\,1~,
$$
for $q$ large enough, namely the unstable bundle is even $\CC^{2}$ (we
only require $\CC^{1+\alpha}$ for some $\alpha>0$). 

The stable bundle is obviously infinitely regular but we can check the
bunching condition for the inverse. We obtain $\lambda=q^{-1}+\Oun
r^{-1}q^{-2}$, $\mu^{-1}=r$, $\nu^{-1}=p+\Oun$. We get
$$
\lambda\mu^{-1}\nu^{-\alpha}\le q^{-1}r p^{\alpha}(1+\Oun q^{-1})\,=\,
\pi q^{-2}p^{\alpha}(1+\Oun q^{-1})~,
$$
and this is smaller than one for $q$ large enough if $\alpha<2$ and $p$
is not much larger than $q$. In other words, we can construct examples
with the stable bundle $\CC^{1+\sigma}$ for any $0<\sigma<1$. 

Finally we have to check the condition $\inf \eta/\varphi>1$. In the above
example this is made simpler by the observation that the set of tangent
vectors with first coordinate equal to zero is covariant. The same is
true for the set of vectors with second coordinate equal to
zero. Therefore, the two invariant bundles are graphs. The largest one is
a set of vectors
$$
\left\{\big(s,0,u_{1}(P)s,v_{1}(P)s,w_{1}(P)s\big)\,\big|\,s\in{\bf R}\right\}
$$
and the lowest one
$$
\left\{\big(0,s,u_{2}(P)s,v_{2}(P)s,w_{2}(P)s\big)\,\big|\,s\in{\bf
R}\right\}~. 
$$
The six functions $u_{i}$, $v_{i}$, $w_{i}$ satisfy the usual coherence
equations, and it follows easily that they are all uniformly bounded by
$\Oun q^{-1}$. It follows easily that 
$$
\eta\ge p+\Oun pq^{-1}
\text{\ \   and\ \   } 
\varphi\le q+\Oun ~,
$$
and our condition is satisfied if $p/q>1+\Oun q^{-1}$ and $q$ is large
enough.
\SECTIONNONR{Appendix}We give here the proof of \clm{petit}. 
The l.h.s.~of \equ{rhoess2} is $\rhoess$. The r.h.s.~will be called
$\bar\rhoess$. 
We have obviously $\bar\rhoess\ge
\rhoess$. To prove the converse inequality, let $\epsilon>0$. From the
definition of $\rhoess$ we can find two subspaces $M$ and $M'$ of
finite codimension such that
$$
\limsup_{n\to\infty}\left(
\sup_{ \scriptstyle x\in M\atop \scriptstyle y\in M'}
{\big|<y,U^{n}x>\big|\over\|y\|\|x\|}\right)^{1/n}< 
\rhoess+{\epsilon\over 3}\;.
$$
This implies that there is an integer $N$ such that for any $n>N$ we have
$$
\left(
\sup_{ \scriptstyle x\in M\atop \scriptstyle y\in M'}
{\big|<y,U^{n}x>\big|\over\|y\|\|x\|}\right)^{1/n}< 
\rhoess +{\epsilon\over 2}\;,
$$
which is equivalent to
$$
\sup_{ \scriptstyle x\in M\atop \scriptstyle y\in M'}
{\big|<y,U^{n}x>\big|\over\|y\|\|x\|}< \left(
\rhoess +{\epsilon\over 2}\right)^{n}\;.
$$
Observe that the spaces $\bar M$ and $\bar M'$ which are the closures of
$M$ and $M'$ are also of finite codimension.
Moreover, for
each $n$, we can find $x\in\bar M$ and $y\in\bar M'$, both of norm 1
 such that
$$
{\big|<y,U^{n}x>\big|\over\|y\|\|x\|}\ge 
\sup_{ \scriptstyle x\in \bar M\atop \scriptstyle y\in\bar M'}
{\big|<y,U^{n}x>\big|\over\|y\|\|x\|}-\left(
\big(\rhoess +\epsilon\big)^{n}-
\big(\rhoess +\epsilon/2\big)^{n}\right)\;.
$$
If $x\notin M$ or $y\notin M'$ or both, we can find two sequences
$(x_{j})\subset M$  and $(y_{j})\subset M'$ converging to $x$ and $y$
respectively. Therefore,
$$
{\big|<y,U^{n}x>\big|\over\|y\|\|x\|}=\lim_{j\to\infty}
{\big|<y_{j},U^{n}x_{j}>\big|\over\|y_{j}\|\|x_{j}\|}\le
 \sup_{ \scriptstyle x\in M\atop \scriptstyle y\in M'}
{\big|<y,U^{n}x>\big|\over\|y\|\|x\|}\;.
$$  
This implies for any $n>N$
$$
 \sup_{ \scriptstyle x\in \bar M\atop \scriptstyle y\in \bar M'}
{\big|<y,U^{n}x>\big|\over\|y\|\|x\|}\le 
\big(\rhoess +\epsilon\big)^{n}~.
$$
Since $\epsilon $ is arbitrary, the result follows.

\LIKEREMARK{Acknowledgments}{We thank V. Baladi and D. Ruelle for
essential help in guiding us through the literature. We have also
profited from helpful discussions with Ch.~Bonatti, M. Viana, and
A. Wilkinson. This 
work was partially supported by the Fonds National Suisse.}

\SECTIONNONR{References}

\eightpoint{
\frenchspacing\setitemindent{[77]}
\def\Rom#1{\uppercase\expandafter{\romannumeral #1}}

}

\begin{thebibliography}{10}

\bibitem{Baladi:1989}
V.~Baladi.
\newblock Unpublished (1989).

\bibitem{Baladi2000}
V.~Baladi.
\newblock {\em Positive transfer operators and decay of correlations\/},
  volume~16 of {\em Advanced Series in Nonlinear Dynamics\/} (River Edge, NJ:
  World Scientific Publishing Co. Inc., 2000).

\bibitem{Bowen1975}
R.~Bowen.
\newblock {E}quilibrium {S}tates and the {E}rgodic {T}heory of {A}nosov
  {D}iffeomorphisms.
\newblock {\em Springer Lect. Notes in Math.\/} {\bf 470} (1975),
  Berlin--Heid\discretionary{--}{}{}elberg--New York.

\bibitem{ColletIsola1991}
P.~Collet and S.~Isola.
\newblock {O}n the essential spectrum of the transfer operator for expanding
  {M}arkov maps.
\newblock {\em Comm. Math. Phys.\/} {\bf 139} (1991), 551--557.

\bibitem{EE}
D.~E. Edmunds and W.~D. Evans.
\newblock {\em Spectral Theory and Differential Operators\/} ({New York}:
  {Oxford University Press}, 1987).

\bibitem{GundlachLatushkin:2001}
V.~Gundlach and Y.~Latushkin.
\newblock {A} sharp formula for the essential spectral radius of {R}uelle's
  operator on smooth and {H}\"older spaces.
\newblock {\em Ergodic Theory Dynam. Systems\/} {\bf 23} (2003), 175--191.

\bibitem{HPS1977}
M.~Hirsch, C.~Pugh, and M.~Shub.
\newblock {I}nvariant manifolds.
\newblock {\em Springer Lect. Notes in Math.\/} {\bf 583} (1977), Berlin--New
  York.

\bibitem{Kato1984}
T.~Kato.
\newblock {\em {P}erturbation {T}heory for {L}inear {O}perators\/}
  (Springer-Verlag, Berlin, 1984).
\newblock Second corrected printing of the second edition.

\bibitem{KatokHasselblatt1995}
A.~Katok and B.~Hasselblatt.
\newblock {\em Introduction to the modern theory of dynamical systems\/},
  volume~54 of {\em Encyclopedia of Mathematics and its Applications\/}
  (Cambridge: Cambridge University Press, 1995).
\newblock With a supplementary chapter by Katok and Leonardo Mendoza.

\bibitem{Keller1979}
G.~Keller.
\newblock {E}rgodicit\'e et mesures invariantes pour les transformations
  dilatantes par mor\discretionary{-}{}{}ceaux d'une r\'egion born\'ee du plan.
\newblock {\em C.R. Acad. Sci. Paris S\'er. A-B.\/} {\bf 289} (1979),
  A625--A627.

\bibitem{Keller1984}
G.~Keller.
\newblock On the rate of convergence to equilibrium in one-dimensional systems.
\newblock {\em Comm. Math. Phys.\/} {\bf 96} (1984), 181--193.

\bibitem{Mattila1995}
P.~Mattila.
\newblock {\em Geometry of sets and measures in {E}uclidean spaces\/},
  volume~44 of {\em Cambridge Studies in Advanced Mathematics\/} (Cambridge:
  Cambridge University Press, 1995).
\newblock Fractals and rectifiability.

\bibitem{Rousseau-Egele1983}
J.~Rousseau-Egele.
\newblock Un th\'eor\`eme de la limite locale pour une classe de
  transformations dilatantes et monotones par morceaux.
\newblock {\em Ann. Probab.\/} {\bf 11} (1983), 772--788.

\bibitem{Ruelle1978}
D.~Ruelle.
\newblock {\em {T}hermodynamic {F}ormalism\/} (Addison Wesley, Reading MA,
  1978).

\bibitem{Ruelle1986PRL}
D.~Ruelle.
\newblock Resonances of chaotic dynamical systems.
\newblock {\em Phys. Rev. Lett.\/} {\bf 56} (1986), 405--407.

\bibitem{Ruelle1987}
D.~Ruelle.
\newblock Resonances for {A}xiom {${\bf A}$} flows.
\newblock {\em J. Differential Geom.\/} {\bf 25} (1987), 99--116.

\bibitem{Young1999}
L.-S. Young.
\newblock {R}ecurrence times and rates of mixing.
\newblock {\em Israel. J. Math.\/} {\bf 110} (1999), 153--188.

\end{thebibliography}
\bye